\documentclass[sigconf]{acmart}

\renewcommand\footnotetextcopyrightpermission[1]{} 
\PassOptionsToPackage{hyphens}{url}\usepackage{hyperref}
\usepackage{breakurl}

\usepackage[nameinlink, noabbrev, capitalize]{cleveref}
\usepackage{soul}

\begin{document}

\title{GAIDE: A Framework for Using Generative AI to Assist in Course Content Development}
\date{August 2023}

\author{Ethan Dickey}
\authornote{Both authors contributed equally to this research.}
\affiliation{%
  \institution{Purdue University}
  \streetaddress{Department of Computer Science}
  \city{West Lafayette}
  \state{Indiana}
  \country{USA}
  \postcode{47907}
  }
\email{dickeye@purdue.edu}
\orcid{0009-0007-3706-5253}

\author{Andres Bejarano}
\authornotemark[1]
\affiliation{%
  \institution{Purdue University}
  \streetaddress{Department of Computer Science}
  \city{West Lafayette}
  \state{Indiana}
  \country{USA}
  \postcode{47907}
  }
\email{abejara@purdue.edu}
\orcid{0000-0003-2611-2855}

\renewcommand{\shortauthors}{Dickey and Bejarano}

\begin{abstract}
    \textit{Contribution}: This research-to-practice full paper presents ``GAIDE: Generative AI for Instructional Development and Education,'' introducing a pragmatic and systematic framework for employing Generative AI (GenAI) in the development of educational content. Unlike existing frameworks, GAIDE emphasizes practical applicability for educators, facilitating the creation of diverse, engaging, and academically sound materials. The novel aspect of our approach lies in its detailed methodology for integrating GenAI into curriculum design processes, thereby reducing instructors' workload and improving the quality of educational materials. Through GAIDE, we contribute a distinct, adaptable model for leveraging technological advancements in education, providing a foundational step towards more efficient and effective instructional material development.
    
    \textit{Background}: The motivation for our study emerges from the increasing demand for innovative and engaging educational content, coupled with the notable rise in Generative AI (GenAI) utilization among students for academic tasks. Our investigations reveal that nearly half of students engage with GenAI tools for completing homework assignments, highlighting a significant shift in study behaviors and the potential for technology to shape educational practices. This scenario presents a dual challenge for educators: to adapt to and incorporate these emerging technologies into their teaching methodologies, not merely to keep pace with technological advancements but to leverage them in fostering a more dynamic and inclusive learning environment. This research addresses these challenges by offering a concrete, adaptable solution, aiming to reshape the landscape of educational content creation and its application across diverse learning settings.
    
    \textit{Intended Outcomes}: The primary objectives of introducing GAIDE are to: 1) Streamline the course content development process for educators, 2) Foster the creation of dynamic, engaging, and varied educational materials, and 3) Demonstrate the practical utility of GenAI in enhancing instructional design, potentially setting a precedent for its adoption in diverse educational contexts.
    
    \textit{Application Design}: GAIDE was conceived out of a necessity to efficiently harness GenAI's potential in education. The application design is rooted in constructivist learning theory and TPCK, emphasizing the importance of integrating technology in a manner that complements pedagogical goals and content knowledge. Our Outcomes-Based Course Design approach aids educators in crafting effective GenAI prompts and guides them through interactions with GenAI tools, both of which are critical for generating high-quality, contextually appropriate content.
    
    \textit{Findings}: Preliminary evaluation of GAIDE indicates its effectiveness in mitigating the instructional challenges associated with content creation. Educators reported a significant reduction in the time and effort required to develop course materials, without compromising on the breadth or depth of the content. Moreover, the use of GenAI has shown promise in deterring conventional cheating methods, suggesting a positive impact on academic integrity and student engagement.
\end{abstract}


\maketitle
\pagestyle{plain}

\tableofcontents

\section{Introduction}
In an era where academic integrity is challenged by the widespread availability of unauthorized solutions online, it has become increasingly important for instructors to generate novel and diverse content each semester. However, in the past year, even freshly created content has come under threat from Generative AI (GenAI) models, which purport advanced language comprehension and question-answering capabilities. While opinions on the actual problem-solving capacities of these models vary, as evidenced by both positive \cite{ouh2023chatgpt, SavelkaEtAl23, Wermelinger23} and negative \cite{kabir23who} reports, there is a consensus among instructors globally about GenAI's potential to significantly disrupt academia, especially in the realm of Computer Science (CS) \cite{lau2023BanIt, Becker2023programmingIsHard, Finnie22, Moradi23}.

A recent preliminary anonymous survey of our summer students enrolled in a data structures and algorithms course revealed that at least 48.5\% utilized GenAI for homework assistance \cite{AILab23}. This figure might even be an underestimation, considering potential hesitations in self-reporting. Recent literature and news articles suggest that the actual number of students using GenAI and the variety of their methods might be more extensive than previously assumed by educators \cite{Yilmaz23, Terry23}. Anecdotally, many students appear comfortable sharing insights about their peers' frequent use of GenAI. Their detailed knowledge of these tools suggests familiarity, although direct admissions of personal use remain rare. All of these factors underscore the growing influence of GenAI in the academic realm but also hint at its potential applications for educators. As the academic community grapples with the ever-growing demand for fresh, diverse, and high-quality course content, the traditional methods of content creation often fall short, being both time-consuming and occasionally monotonous.

While the challenges posed by GenAI to academic integrity are evident, it is crucial to recognize the transformative potential these tools offer. The same capabilities that enable students to seek unauthorized assistance can, if channeled correctly, revolutionize the way educators create and deliver content. Content Creators (CCs), in particular, find themselves uniquely positioned in this landscape: they stand to benefit directly from GenAI's advantages in content generation, such as creating diverse, high-quality, and relevant content, yet they also confront the challenges it presents to academic integrity. This dual impact places CCs at a critical juncture. Without a structured approach, they risk either not fully harnessing the benefits of GenAI or inadvertently amplifying its challenges \cite{Zamfirescu2023nonExpertsFailPrompts, white2023promptpattern, lau2023BanIt}. Therefore, a systematic method is essential to guide CCs in navigating the complexities of GenAI, ensuring they can effectively leverage its benefits while being aware of its potential pitfalls. This burgeoning potential brings forth a pivotal question: \textbf{how can educators harness the power of GenAI to create meaningful course content efficiently?} It's worth noting that the challenge of unauthorized student assistance with GenAI is a significant concern and is addressed in detail in our separate parallel work \cite{AILab23}.

In light of the growing utility and student interest in GenAI tools, this paper aims to develop a generalized approach for CCs in academia to harness the potential of these tools. Specifically, we study natural language GenAI models which incorporate memory of conversation (the authors tested the framework using ChatGPT 3.5 and 4.0\footnote{\url{https://chat.openai.com/}}, Bard\footnote{\url{https://bard.google.com/}}, Llamma\footnote{\url{https://ai.meta.com/llama}}, and Microsoft Bing's Copilot\footnote{\url{https://www.bing.com/}} and found ChatGPT 4.0 to produce the best overall results with our framework as of November 2023). We begin by substantiating our observations on GenAI's utility and student interest, providing both rationale and illustrative examples. After motivating why instructors should be interested in mastering these tools, we introduce the \textit{GenAI Content Generation Framework}. Remarkably, up until the submission date of this paper, we have encountered no preexisting framework specifically aiding educators in incorporating GenAI tools within the content creation process. Most existing literature often adopts a broader academic perspective, emphasizing empowering students to critically assess these tools and fostering ethical conversations \cite{Bull23frame, Gutierrez23, Guerriero23, Su23}. In contrast, our approach stands out by specifying a particular workflow and offering practical recommendations tailored for CCs. Furthermore, we provide an explicit rationale for its adoption, ensuring that CCs not only understand the 'how' but also the `why' behind each step. This framework, characterized by a flow of steps and guiding perspectives, assists CCs in harnessing GenAI efficiently to systematically and practically get high-quality results. While each interaction with GenAI models is unique due to its dynamism and creativity (and pseudo-random hallucinations), our framework serves as a consistent guide, offering practical strategies to achieve precise content outcomes. Subsequent sections delve into broader considerations for engaging with GenAI and conclude with reflections on the framework's implications and potential future research directions.

\section{Why Should Educators Consider GenAI for Course Content Creation?}
\label{sec:why}
GenAI excels at assisting experts in executing simple tasks more efficiently and, in many cases, with enhanced outcomes, as detailed in subsequent subsections. While experts can harness the power of GenAI effectively, non-experts (such as students) face distinct challenges \cite{Lee2023}. They may struggle to verify, correct, and selectively use results, leading to potential pitfalls such as the reinforcement of incorrect knowledge. Other significant risks for non-experts include a lack of genuine skill development, diminished problem-solving capabilities, and an over-reliance on the tool, which can hinder self-sufficiency.

The advent of Large Language Models (LLMs), such as OpenAI Codex, marked a turning point in CS education, a trend that gained considerable momentum with the GenAI explosion of 2022 and 2023. Pioneering studies, like those by Sarsa \textit{et al.} \cite{Sarsa2022}, explored using these models to craft programming exercises, solutions, and test cases. These early experiments demonstrated the potential for LLMs to generate innovative and relevant educational content, albeit requiring meticulous instructor oversight and adjustments. This initial exploration laid the groundwork for today's educators, who, with advanced LLM capabilities and more refined techniques in prompt engineering, can now much more efficiently and effectively create and tailor course materials. This evolution reflects a growing sophistication in the application of GenAI tools in education, setting the stage for their broader utilization, as detailed in the subsequent sections.

Recent studies and observations have underscored the multifaceted capabilities of GenAI \cite{lau2023BanIt, white2023promptpattern, Becker2023programmingIsHard, Buriak2023best, Tlili2023devil, Shidiq2023creative}. These models excel in generating large volumes of content swiftly, drafting documents, automating repetitive interactions, providing interactive explanations, generating and evaluating code, and producing (arguably) creative content. For course CCs, these capabilities translate into tangible benefits. They can produce content more rapidly, brainstorm and ideate with the assistance of the model, and refine content iteratively based on conversational or formal feedback. The direct advantages of harnessing these properties include faster course development, increased content flexibility (styles, levels, etc.), the ability to update content frequently, scalability of personalized content, cost-effective content creation, support for experimental approaches, and automation of repetitive tasks.

While many of these benefits provide intrinsic motivation for course CCs to explore these tools, we also present a student-centered perspective. We posit that (a) students are highly likely to experiment with these tools, and (b) these tools are here to stay. To truly understand the implications of these tools within academia, instructors would benefit from firsthand experience with them.

To substantiate point (a), we highlight several observations. First, the detection of GenAI tool usage for academic dishonesty has proven challenging, given its capacity to produce seemingly original content (including craftily rephrasing and editing) \cite{weberwulff2023testing, DALALAH2023100822}. This makes it arguably more elusive than traditional forms of academic dishonesty. Second, conventional deterrents against academic dishonesty face challenges in the context of GenAI due to its novelty and accessibility. Unlike instances where students source content from the internet, GenAI-generated content is unique, making detection and prevention more complex. Lastly, there's a growing consensus that these tools will be permissible and even prevalent in future workplaces, a sentiment echoed by CS instructors globally \cite{lau2023BanIt}. We can see these conversations happening even now, from legal perspectives \cite{Grossenbacher2023, Bates_2023} to strong corporate stances on both sides of the line \cite{Telford_Verma_2023, DeRose_2023}. Given this trajectory, students often pose a challenging question: ``Why \textit{shouldn't} I use GenAI?'' We delve deeper into this dilemma in a separate study, where we introduce a method to address this very concern \cite{AILab23}. In this paper, we contend that the rapid adoption of GenAI by students and industries underscores the urgency for educators and course CCs to fully grasp its academic ramifications. This ensures they remain at the cutting edge of this dynamic academic environment. Presently, the most effective method to achieve this comprehension is through direct engagement and application of these tools.

To support (b), we share a quote from a recent investigation into GenAI by McKinsey \cite{McKinsey2023}, one of the oldest and largest of the world's most prestigious management strategy consulting firms, which we believe aptly summarizes current views on the future of GenAI: \textit{``All of us are at the beginning of a journey to understand [GenAI’s] power, reach, and capabilities... [This research] suggests that [GenAI] is poised to transform roles and boost performance across functions such as sales and marketing, customer operations, and software development. In the process, it could unlock trillions of dollars in value across sectors from banking to life sciences.''}

In light of the above discussions and the insights from McKinsey, it becomes evident that GenAI is not just a fleeting technological trend but a transformative force poised to reshape various sectors, including academia. As educators and stakeholders in the academic community, it is incumbent upon us to not only recognize the challenges posed by GenAI but also to proactively engage with it. By doing so, we can harness its potential for positive educational outcomes while mitigating risks. This proactive approach will ensure that academia remains adaptive, relevant, and prepared for the evolving landscape that GenAI presents. As we navigate this new frontier, collaboration, continuous research, and open dialogue will be paramount in guiding our path forward.

\section{GAIDE: A GenAI Content Generation Framework}
\label{sec:framework}
Properly motivated, we dive into the foundational principles and structure of our GenAI Content Generation Framework. This framework is most succinctly described as a sequence of steps, each accompanied by its respective perspective, guiding the integration of GenAI in collegiate-level course content development. While our approach mirrors traditional content development processes, we have specifically aligned it with the Outcomes-Based Course Design methodology \cite{harden1999OutcomesBased}. This choice is influenced by Ziegenfuss's summary of observed approaches to course design \cite{ziegenfuss2007phenomenographic} and the widely recognized Backward Course Design Model \cite{Hansen_2011}.

Building further on foundational methodologies, the GAIDE framework is deeply embedded within constructivist learning theory \cite{gilakjani2013teachers}, which posits that learners construct knowledge through active engagement with their environment. This aligns with our aim to employ GenAI to create dynamic and interactive learning experiences that are custom-tailored to meet the diverse needs of students. Moreover, we integrate the principles of Technological Pedagogical Content Knowledge (TPCK) \cite{koehler2009technological}, ensuring that technology through GenAI supports and actively enhances pedagogical practices and content delivery. This strategic alignment guarantees that GAIDE supports educational objectives while also advancing the practical application of these theories in real-world educational settings.

By rooting the GAIDE framework in these diverse theories, we provide a structured approach to content generation and enrich the ongoing discourse on integrating technology in education. This theoretical grounding ensures that our framework not only streamlines educational processes but actively enhances them, creating an environment where educators and students alike can thrive. The application of constructivist and TPCK principles underscores a proactive approach to improving educational practices, ensuring that technology serves as a bridge rather than a barrier to effective learning."

\begin{figure*}
    \centering 
    \includegraphics[scale=0.55]{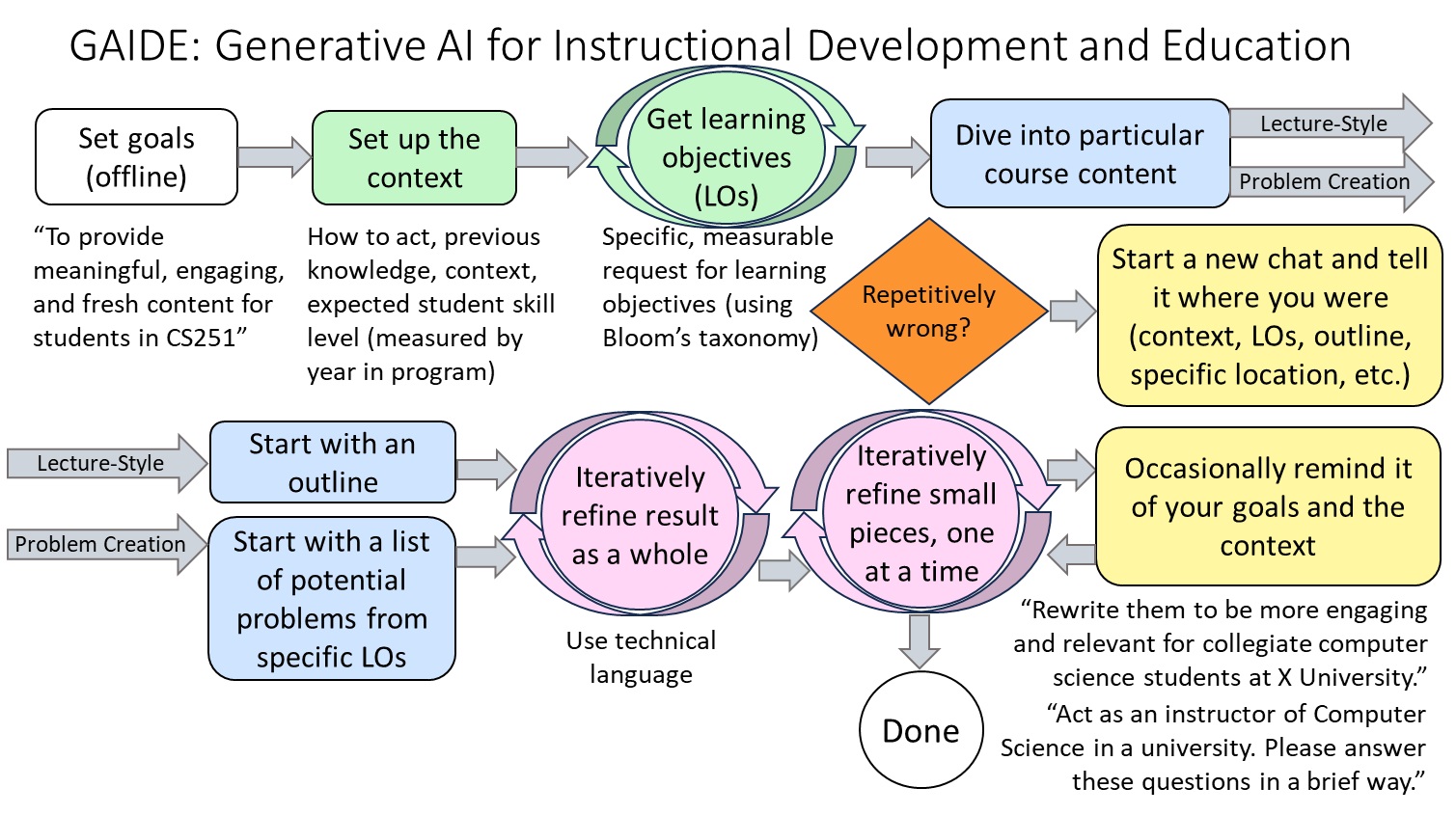}
    \caption{A illustrative overview of the GAIDE process. Please refer to Section \ref{sec:framework} for further discussion of each component. Please refer to the Supplementary Materials (available below main text body) for a simple example of the framework using ChatGPT4.0 and Microsoft Bing's Copilot}
    \label{fig_gaide}
\end{figure*}

After designing course outcomes, the framework moves to course content draft generation. This is followed by iterative refinement, initially on a macro-scale and subsequently on a micro-scale. Recognizing the diverse nature of course content and the versatile capabilities of GenAI, we've categorized our approach into two primary content types: Lecture-Style and Problem Creation. While our framework doesn't adhere strictly to a specific design model, it offers flexibility and adaptability to various educational contexts. We wrap up this section with additional recommendations for harnessing GenAI, which, while valuable, didn't seamlessly fit within the main structure of the framework. Figure \ref{fig_gaide} contains a visual illustration of the following process, and the Supplementary Materials (available below main text body) contain a simple example of the framework using ChatGPT 4.0 and Microsoft Bing's Copilot.


It is important to note that our framework primarily targets the creation of course content for undergraduate collegiate CS courses. Nevertheless, the versatility of this framework allows for its adaptation to a variety of educational contexts. As course material delves deeper and becomes more advanced, there's an increased likelihood of GenAI models producing inaccurate or misleading content. Despite these potential pitfalls, GenAI models remain invaluable tools for content design and ideation, even in advanced courses.

\subsection{Setup}

\subsubsection{Set Goals}
The first and only offline (without GenAI) step in our framework is the establishment of clear goals. These goals not only guide the narrative presented to the GenAI model but also serve as a constant reminder of the intended direction. For instance, when we were designing new homework and tests for a summer class, one of our primary goals was: ``to provide meaningful, engaging, and fresh content for students in CS251.'' While these goals can be as broad or specific as desired, it is imperative that they contain specific, measurable outcomes. To clarify, a measurable outcome might be something like ``students should be able to implement a basic sorting algorithm'' rather than a vague goal like ``students should understand sorting.'' In the process of setting and working towards these goals, it's also worth noting the importance of self-reflection. Challenging one's own assumptions about what these models can achieve can be beneficial, as it fosters adaptability and encourages a proactive approach to potential limitations and innovations of GenAI. By regularly reassessing and adjusting one's expectations, educators can better harness the evolving capabilities of these models. For our team, the challenge lay in the ``fresh content'' aspect. Given the numerous semesters of content we had accumulated, devising non-repetitive material had proven to be a significant hurdle. Contrary to our initial assumptions, we found that these tools were particularly adept at generating fresh content, often surpassing our own initial expectations.

\subsubsection{Set Up the Context}
Transitioning to the online component of our framework, the first essential step with any GenAI session is to set up the context. Broadly, this involves providing the model with details akin to what a new DTA\footnote{We refer to TAs who develop content for courses as one of development TAs, dev TAs, or DTAs.} would require to design the course content. Specifically, to get quality results, one should tell the model how to act (known as the `persona prompt pattern' \cite{white2023promptpattern}), the topic under consideration, any prior knowledge the students possess, and any pertinent student demographics (usually just their academic year). For example, high schoolers in a collegiate summer program possess a different skill set and background compared to sophomore CS majors. Providing the model with such demographic details enables it to tailor content with greater precision (additional benefits surrounding demographics and perspectives are discussed in \ref{observation_perspectives}). Our approach to context setup aligns with the `context manager prompt pattern' introduced by White \textit{et al.} \cite{white2023promptpattern}. However, our framework offers specific recommendations on context details and their placement within the GenAI session.

\subsubsection{Generate Learning Objectives}
Concluding the setup stage, the generation of learning objectives (LOs) is paramount. The centrality of LOs in guiding subsequent interactions with GenAI cannot be overstated. This emphasis on LOs is rooted in our adoption of the Outcomes-Based Course Design methodology \cite{harden1999OutcomesBased, ziegenfuss2007phenomenographic}, which prioritizes determining student outcomes from the outset. For a deeper exploration of LOs as foundational to effective teaching, refer to \cite{felder2016teaching}.

When instructing GenAI to generate LOs, it's essential to make specific, measurable requests. Additionally, directing the model to employ professional language, such as terms from Bloom's revised taxonomy \cite{revisedBloomsTaxonomy}, is crucial. This not only implies a quality standard for the LOs based on other LOs which use the same language but also communicates the desired specificity and formality to the model. Below are sample prompts to guide GenAI: \textit{``Act as an \ul{instructor of Computer Science in a university}. The current topic we are covering is \ul{Binary Search Trees}. Students know \ul{discrete math, basic programming, pointers, primitive data structures, and runtime algorithm analysis}. Students are in their \ul{second} year of studies. Give me \ul{five} LOs for the current topic. Use Bloom's revised taxonomy verbs for the objectives.''}


As we delve into design sessions with GenAI, it is important to grasp how best to interact with the model. The generation of content, including LOs, is iterative, with initial outputs often not aligning perfectly with the specific course goals. This is a key reason for requesting a larger number of LOs than might be immediately needed. From this broader set, educators can select a subset that resonates more closely with their goals and the targeted student profile. If the initial LOs don't fully meet expectations, it is beneficial to work with this chosen subset, providing the model with targeted feedback for refinement. For instance, if an objective needs rewording to better fit within Bloom's revised taxonomy or to emphasize application over theory, such iterative feedback can steer the model accordingly. An illustrative interaction might be: ``I like learning objectives 2, 3, and 7, but I want you to reword 2 to be higher in the Cognitive Process dimension of Bloom's revised taxonomy and adjust 3 to prioritize application over theoretical understanding.'' Lastly, if you edit something the model gave you, it is imperative to let it know what you did, even if you do not want its feedback. This allows it to learn your preferences over a session.

\subsection{Course Content Rough Draft}
Once refined LOs are established, the focus shifts to the specific course content. While course content can take various forms, for the purposes of this framework, it is categorized into two primary types: Lecture-Style and Problem Creation. A rough draft of the former refers to an outline that should be correct only from a very high-level view, while a rough draft of the latter refers to lists of potential problems, from which only a select subset will be chosen for iterative refinement. Following the creation of these rough drafts, we move to iterative refinement from a macro perspective.

\subsection{Macro Refinement}

It is critical to understand the distinction we make between macro and micro refinement. Macro refinement focuses on adjusting the content draft holistically, refining entire sections to ensure alignment with the established expectations, goals, LOs, and context. Conversely, micro refinement zeroes in on specific parts of the draft, addressing them with precision and individualized context. At the macro stage, it is acceptable if certain components don't fully meet expectations, as long as they are broadly correct; finer adjustments are reserved for the micro refinement phase.

During macro refinement, interactions with the model should mirror discussions with a dev TA\footnotemark[5], employing technical terminology and offering feedback on the draft's strengths and weaknesses.

With this foundation, we present considerations for each category and provide guidance on when to transition to micro refinement.

\subsubsection{Lecture-Type}

In any content type which requires a larger flow and organization, outlines are typically a good starting place. Request an outline appropriate for your content and specify which LOs you wish to use in it. From there, you should iteratively refine the outline as a whole. Here are some things to consider:
\begin{itemize}
    \item \textbf{Duration:} How long is the lecture?
    \item \textbf{Associated Tasks:} Are there any associated tasks outside the lecture, such as grading in-class activities or preparing materials?
    \item \textbf{Pre- and Post-Lecture Activities:} Are there any activities before or after the lecture?
    \item \textbf{Subtopics:} Are there any subtopics that are crucial to cover?
    \item \textbf{Specific Activities}: Are there particular activities, such as quizzes or group work, that are planned?
\end{itemize}

Each of these considerations, if relevant, should be communicated to the model during the refinement process (e.g. \textit{``Rewrite the outline to fit into 50 minutes''}). Addressing all considerations simultaneously or tackling them individually are both valid approaches, with both methods showing comparable success rates in our experience (e.g. \textit{``Try again, but make them all higher in Bloom's Taxonomy''} versus \textit{``I like 2 and 3, but 1 and 4 don't match with my goals, please make 1 harder and 4 easier, to make this more appropriate for a timed exam setting''}).

Additionally, it can be helpful and incredibly insightful to let GenAI brainstorm on different lecture components. Moreover, encouraging the model to deviate from routine approaches aligns with the active learning principle of unpredictability \cite{felder2016teaching}. For a deeper dive into brainstorming and alternate perspectives, refer to \ref{observation_perspectives}.

Once the outline aligns with the lecture's objectives and approach, it's time to delve into specific sections, similar to the process for problem sets. At this juncture, managing expectations is crucial. The model might produce a near-perfect outline or offer just a few valuable section ideas. Regardless, these outputs serve as a foundation for the micro refinement phase, where the lecture content can be further tailored for a complete draft or cherry-picked for integration into an existing draft.

\subsubsection{Problem Creation}
For content types that result in lists, such as problems or activities, the emphasis is on the diversity of responses and the general applicability of the items, hereafter referred to generically as ``problems.'' Unlike Lecture-Type content, only a subset of the generated list will be included in the final version. During the micro refinement phase, the focus will be on a select few top problems rather than an exhaustive review of every item as is done for outlines.

To generate a draft, ask the model to generate a specific number of problems, ideally more than required, based on the selected LOs. If the generated problems do not align with the intended goals and context, provide feedback to the model, emphasizing the desired attributes. Key variables to consider include:
\begin{itemize}
    \item \textbf{Answer style:} Multiple choice, short answer, etc.
    \item \textbf{Depth:} Desired level(s) within Bloom's Revised Taxonomy \cite{revisedBloomsTaxonomy} or Webb's Depth of Knowledge \cite{depthOfKnowledgeWebb}.
    \item \textbf{Theme:} Leveraging GenAI's strength in creativity.
    \item \textbf{Topical theme:} Ensuring the problems address specific skills within the topic.
\end{itemize}

The macro refinement process can vary based on the alignment of the generated problems with the goals. For instance, if only a few questions are relevant, instruct the model to generate problems similar to the relevant ones. If the model's responses become repetitive or misaligned, providing a sample problem can be particularly helpful in resetting its misconceptions about what is desired. Several other examples exist, but we leave it to the discretion of the reader to respond to the model within the spirit of this framework.

The boundary between macro and micro refinement in Problem Creation can be nebulous. A good indication of the transition to micro refinement occurs when the focus shifts to refining a select subset of problems, while disregarding the rest.

\subsection{Micro Refinement}
Before delving into micro refinement, educators should be satisfied with the overall structure and general alignment of the content draft with the intended goals and context. From there, the focus shifts to perfecting and elaborating on each segment of the content. This involves examining each section, subsection, or question individually, informing the model of the specific focus on that part.

During these small, focused refinements, it is crucial to maintain context specificity. For instance, if refining a problem on frequency analysis in Huffman Coding emphasizes the theoretical aspect over the practical skill, this context should not inadvertently influence a subsequent question on creating the Huffman Coding Tree, where hands-on application is integral to understanding the theory. See \ref{sec:framework:contextintegrity} for further discussion on context integrity.

Alongside specifying the content segment for refinement, provide the model with clear modification instructions. The granularity of these requests can vary widely. For instance, feedback can range from broader directives like, \textit{``make this question more challenging; it's currently too straightforward for a homework assignment,''} to more specific directives such as, \textit{``I find the word `target' unsuitable; could you rephrase that sentence, please?''} Such detailed interactions ensure the content aligns closely with the educator's vision and objectives.

The micro refinement process is inherently flexible, adapting to the educator's vision for the final content. While there's no singular correct approach, we offer two potential workflows for each content type.

\subsubsection{Lecture-Type}

Once satisfied with a particular segment of the outline, educators can request the model to generate a detailed script or essential talking points. These scripts serve as a roadmap for the lecture, ensuring a coherent flow and comprehensive coverage of the topic. As the script is generated, it's crucial to assess its alignment with the LOs and its potential to engage students. Any misaligned or inaccurate sections should be refined iteratively. Once sections are polished, the model can merge the script cohesively. This refined script can be paired with lecture slides, multimedia, or classroom activities to enrich the educational journey.

\subsubsection{Problem Creation}
In the refinement of individual problems, several strategies emerged as particularly effective. Among these, iterative rewording of the problem by the model, adjusting the problem's difficulty, and embedding problems within a narrative stood out. Notably, while most narrative integrations are typically part of micro refinement, exceptions arise when a global storyline is employed, where each problem contributes to a larger narrative.

During problem refinement, educators can instruct the model to answer the question from an undergraduate's perspective (or some other demographic(s)). This approach can reveal common pitfalls and misconceptions, guiding further refinement. Such insights are invaluable, especially for problems where students must comprehend and act without instructional support. Misunderstandings can arise in activities without instructional support, often over aspects not intended for assessment. Rewording the problem can clarify these points, enhancing comprehension.

As the refinement progresses, educators should request the correct answer (usually requesting in a brief or concise way to avoid the overly-wordy responses certain models are prone to). If accurate, this stage is suitable for generating a rubric. While not always standard practice, rubrics ensure a consistent and fair assessment. For a deeper understanding of rubric benefits, Ragupathi and Lee \cite{ragupathi2020beyond} provide valuable insights.

If the model's answer is incorrect, educators can guide it towards the right solution. Persistent errors can highlight tasks that challenge the model, incredibly powerful information which offers unique teaching opportunities about the limitations of GenAI (as address in \cite{AILab23}). Regardless, perfection in answers isn't the goal. The quality of a rubric doesn't hinge on the exact correctness of the provided answer, but rather on sound evaluation criteria.

While the iterative refinement process may seem labor-intensive, it's essential to note that these steps are integral to traditional content creation. GenAI often streamlines this process, reducing the time and effort typically required.

\subsection{Maintaining Contextual Integrity in Iterative Refinement}\label{sec:framework:contextintegrity}

In the process of iterative refinement, two critical concerns arise: \textit{context blending} and \textit{loss of focus}. The term \textit{`context blending'} pertains to the merging of contexts between different steps that do not inherently share the same context. On the other hand, \textit{`loss of focus'} denotes the model's diminishing capacity to execute the tasks and refinements as directed. To mitigate context blending, one should delineate clear transitions between sections and intermittently reinforce context-specific labels, such as LOs, sections, or parts under consideration. Furthermore, prompting the model to reiterate these context-specific labels as it transitions to a subsequent section can enhance its adherence to the expected context. Most crucially, if a loss of focus is observed, it may be beneficial to initiate a complete context reset. This can be achieved by starting a new session, reintroducing the context, and specifying the current stage of the process—by providing, for instance, the working outline, LOs, global context, and the specific section in focus.

\subsection{Consolidating Generated Content}

Given the intricacies of generating comprehensive course content, there is a high likelihood that it will be necessary to initiate multiple sessions with GenAI. Due to this segmented approach, the model may not seamlessly provide an overarching summary of all content components in a singular response. As a practical measure, CCs are encouraged to maintain a dedicated document to collate and refine the selected outputs, ensuring a cohesive and well-structured course assembly.

\subsubsection{GenAI in Comprehensive Course Planning}

While the primary focus of this framework is not on constructing an entire course from the ground up using GenAI, the potential of these models in the realm of course planning warrants attention. Once LOs, course outlines, and assessments are established, GenAI models demonstrate a commendable proficiency in devising comprehensive course activity plans within a single session. This includes detailed time allocations and other integral components of a structured plan. Leading chat-based GenAI enterprises, such as OpenAI, Anthropic, and Google, have underscored the prowess of these tools in planning, brainstorming, and feedback solicitation. Notably, the planning capability of GenAI shines when tasked with orchestrating plans for cohesive sets of course content, rather than isolated components.

\section{General Considerations for Engaging with GenAI}
\label{sec:general}
In this section, we offer various insights and recommendations pertinent to interacting with GenAI. These considerations, while valuable, do not align explicitly with any specific segment of the framework.

\begin{enumerate}
     \item \textit{Perspective.} A useful analogy for understanding a GenAI model likens it to a child aged 6-8 years. Such children excel at executing tasks when given clear instructions, provided they possess the requisite knowledge. In the case of GenAI, instead of drawing from 6-8 years of human experiences, it taps into the vast expanse of the internet. However, akin to children of this age, GenAI models often lack initiative and may not perform tasks without explicit direction. Absent specific guidance, the assumptions made by the model could diverge significantly from the intended objectives of the session.
     \item \textit{Brainstorming.} One of GenAI's notable strengths lies in its creative prowess. Rather than providing explicit specifications for assessments or activities, CCs might find it beneficial to solicit recommendations from GenAI for various course content elements. This approach can yield diverse and innovative ideas, enriching the educational experience.
     \item \textit{Embracing Imperfection.} While the pursuit of excellence is commendable, it's essential to recognize the inherent limitations of GenAI. Continuously striving for perfection can lead to diminishing returns. Instead, it's often more productive to acknowledge areas where the model may falter and focus on harnessing its strengths. In essence, avoid getting mired in endless refinements and capitalize on the valuable insights it provides.
     \item \textit{Optimizing Content Generation.} When soliciting GenAI for new content, always request a more extensive set than required. This approach allows for a selection process, ensuring the final subset aligns closely with goals and context. Additionally, by providing specific directives, such as \textit{``with varying levels of difficulty''} or \textit{``answer in a brief way,''} one can enhance the utility and precision of the generated output.
\end{enumerate}

\subsection{Diversity of Perspectives} \label{observation_perspectives}

In the realm of education, understanding and addressing the diverse backgrounds and experiences of learners is paramount. One of the unique capabilities of GenAI is its ability to simulate a multitude of perspectives, which can be instrumental in illuminating potential blind spots in course content, especially when creating content for unfamiliar demographics.

A particularly effective strategy involves employing the phrase \textit{``act as''} when interacting with GenAI (known as the `persona prompt pattern' \cite{white2023promptpattern}). By instructing the model to \textit{``act as''} a particular demographic or adopt a specific perspective, educators can gain insights that might otherwise remain obscured. For instance, asking GenAI to \textit{``act as a student from a non-technical background''} or \textit{``act as an international student''} can yield content and feedback that is more attuned to the needs and challenges of these specific groups.

Leveraging this feature not only enriches the educational material but also fosters an inclusive learning environment. It ensures that content is not inadvertently biased or neglectful of the diverse experiences and backgrounds that students bring to the classroom. In essence, by harnessing the diversity of perspectives that GenAI can offer, educators can craft a more holistic and inclusive educational experience.

\section{Discussion and Future Work}
\label{sec:discussion}
Given the relatively recent release and rise to prominence of easily accessible GenAI tools, research studying their usage and applications is still very young. Zhai \textit{et al.} \cite{Zhai20} concluded that ``[Machine Learning] has transformed--but not yet redefined--conventional science assessment practice in terms of fundamental purpose, the nature of the science assessment, and the relevant assessment challenges.'' While not addressing the purpose and nature of assessments themselves, the presented framework takes one small step in pushing to redefine how traditional course development is done - and seeks to change who is qualified to develop high-quality course content.

Furthermore, this framework provides a novel perspective on what could and should be done with these tools. Instead of asking, ``what should we do about these tools,'' as many of our colleagues and professors around the world appear to be focusing on \cite{lau2023BanIt}, this framework turns the question around on instructors and facilitates discussion surrounding ``how can these tools help \textit{us} in all of our activities?''

While our primary focus in this study was to elucidate the potential of GenAI in assisting educators with content creation, the second research question posed in our abstract remains an essential area of inquiry: Can the use of GenAI significantly reduce the workload of instructional staff? This question is of paramount importance, especially in the context of increasing class sizes and the constant demand for updated course materials. A specific and measurable research question could be formulated as: ``To what extent can the integration of GenAI in course content development reduce the time spent by instructional staff on content creation and revision?'' Future studies should employ both quantitative and qualitative methods to assess the time savings, if any, and the potential shifts in the nature of the workload. For instance, while GenAI might reduce the time spent on content creation, it might introduce new tasks, such as refining GenAI outputs or tailoring generated content to specific course objectives. Addressing this research question will provide a more comprehensive understanding of the true impact of GenAI on the educational landscape.

In an effort to gauge the initial reception and applicability of our framework, we introduced it to around 20 of our colleagues during a series of workshops. Although these workshops were conducted recently, and a comprehensive analysis of the survey results is pending, preliminary feedback suggests a high degree of satisfaction with the framework and an indication that the workshop objectives were largely met. Additionally, having used this framework for nearly a year, we have found it to be frequently beneficial and versatile across a wide variety of course contexts and settings.

While we sought to be as general and extendable as possible, the applicability and utility of this framework are inherently bound to some degree to the selected GenAI model. It works best on a chat-based, highly trained model that understands nuances and perspectives across academia. Furthermore, the development of this framework was performed in the context of CS undergraduate courses. More work will be required to validate its utility in other contexts and potentially adapt the model to one that is more useful across disciplines. The GAIDE framework is designed to be highly adaptable and applicable across disciplines by aligning with the specific learning objectives and goals of each respective field. This flexibility allows GAIDE to support educators in different contexts, ensuring the framework remains relevant and practical regardless of the subject matter or educational level.

To conclude, the authors would like to leave you with two thoughts: First, as the frontiers of technology continue to expand, the essence of education remains deeply rooted in the dynamic interplay between innovation and tradition. In the digital age, where generative AI reshapes learning landscapes, it beckons us to adapt and reimagine the educational paradigms. With the GAIDE framework, we endeavor to harness the formidable potential of generative AI, not as a replacement for the human touch in education but as a complement that enriches it. Let us consider how these tools can be molded to respect and uplift the timeless values of teaching while also preparing learners to thrive in a world where change is the only constant.

Second, in the relentless pursuit of progress, the realm of education stands as both a beneficiary and a steward of technological evolution. As we integrate these advanced tools, let us remember that their true value lies not in their ability to replicate human thought but in their capacity to expand the horizons of what educators can achieve. In this new educational landscape, our challenge is to cultivate a synergy where technology amplifies creativity, enhances inclusivity, and deepens understanding, thereby preparing a generation that is as wise as it is technologically adept.

\begin{acks}
The authors would like to thank Purdue's Center for Instructional Excellence, at whose generative AI discussions we met and were encouraged to pursue our ideas. Furthermore, a special thanks to Emily Bonem and David Nelson for supporting and encouraging us initially and repeatedly as we execute our project.  Lastly, we would like to thank Andre Beasley for a useful discussion on general course development using ChatGPT.
\end{acks}

\bibliographystyle{ACM-Reference-Format}
\bibliography{refs}

\title{Supplementary material: A Framework for Using Generative AI to Assist in Course Content Development}


\begin{center}
    {\LARGE \textbf{Supplementary Material: A Framework for Using Generative AI to Assist in Course Content Development}} \\
    \vspace{1em}
\end{center}


\Cref{sec_chat4} contains a ChatGPT4.0, \Cref{sec_bingco} contains a nearly identical interaction with Bing Copilot.

\section{ChatGPT 4.0}\label{sec_chat4}
This is a nearly identical conversation (from the user's side) with ChatGPT 4.0 as we had with Bing Copilot.

Conversation took place in November of 2023.

\textbf{User: }Act as a professor of computer science from a top 10 engineering university. You are teaching a course about data structures and algorithms to Computer Science and Data Science majors (students take this course in their sophomore year). So far, you have covered primitive data structures, asymptotic runtime analysis, and dynamic data structures. The last class was about binary search trees, emphasizing the problems we could get with them. Please give me five learning objectives for a class discussing red-black trees. Use Bloom’s revised taxonomy of verbs for the objectives.

\textbf{ChatGPT 4:} Certainly! When structuring learning objectives for a class on red-black trees using Bloom's revised taxonomy, we want to ensure that the objectives cover a range of cognitive processes, from remembering and understanding basic concepts to applying, analyzing, evaluating, and creating based on these concepts. Here are five learning objectives for a class on red-black trees:

\begin{enumerate}
    \item \textbf{Remember:} Recall and describe the properties that define a red-black tree, including the color of nodes and balance requirements.
    \item \textbf{Understand:} Explain the significance of each red-black tree property in maintaining a balanced tree and how it affects the tree's performance for operations such as insertion, deletion, and search.
    \item \textbf{Apply:} Implement the insertion operation in a red-black tree, ensuring that the tree maintains its defining properties after adding a new node.
    \item \textbf{Analyze:} Compare and contrast the mechanisms of balancing in red-black trees with those in AVL trees, highlighting the differences in rotations and color changes.
    \item \textbf{Evaluate:} Assess the advantages of using a red-black tree in a particular scenario, such as databases, and justify its suitability over other types of binary search trees.
\end{enumerate}

By the end of the class, students should have a comprehensive understanding of red-black trees, including their structure, operations, and practical applications, as well as an ability to critically evaluate their use in various computing scenarios.

\textbf{User:} I like your first three learning objectives! With these in mind, please propose a lecture outline for a 50-minute class.

\textbf{ChatGPT 4:} Certainly! Here’s a lecture outline for a 50-minute class on Red-Black Trees designed around the first three learning objectives.

\textbf{Lecture Outline: Red-Black Trees}

\begin{enumerate}
    \item \textbf{Introduction (5 minutes)}
    \begin{itemize}
        \item Brief review of binary search trees and their limitations.
        \item Introduction to red-black trees and their importance in maintaining balanced trees.
    \end{itemize}

    \item \textbf{Recap of Red-Black Tree Properties (10 minutes)}
    \begin{itemize}
        \item Objective: Remember the defining properties.
        \item Enumerate the properties of red-black trees.
        \item Show visual examples of each property for clarity.
    \end{itemize}

    \item \textbf{Understanding Red-Black Trees (10 minutes)}
    \begin{itemize}
        \item Objective: Understand the significance of properties.
        \item Explain why red-black trees remain balanced (brief touch on the concept of "black height").
        \item Discuss the impact of red-black tree properties on operation performance.
    \end{itemize}

    \item \textbf{Red-Black Tree Insertion Operation (15 minutes)}
    \begin{itemize}
        \item Objective: Apply the insertion operation.
        \item Step-by-step insertion algorithm explanation.
        \item Live coding example or walk through an animated insertion process.
        \item Discuss the cases of re-coloring and rotations after insertion.
    \end{itemize}

    \item \textbf{Class Practice and Interactive Problem Solving (5 minutes)}
    \begin{itemize}
        \item Students try to insert a node into a red-black tree on paper or using classroom software.
        \item Quick discussion and resolution of common issues or mistakes.
    \end{itemize}

    \item \textbf{Q\&A Session (5 minutes)}
    \begin{itemize}
        \item Address student questions and clarify doubts.
        \item Encourage questions that lead to deeper understanding and application of concepts.
    \end{itemize}
\end{enumerate}

Each section of the lecture is designed to build upon the previous one, ensuring that students have a clear and comprehensive understanding of red-black trees by the end of the session. The interactive problem-solving segment allows students to apply their knowledge immediately, which can help to reinforce learning and identify areas that may need more explanation.

\textbf{User: }Let's review items 4 and 5. They give too much attention to the insert operation and no mention of the delete operation. Please adjust these two items properly so both operations are discussed.

\textbf{ChatGPT 4:} Certainly! Adjusting the outline to give equal attention to both insert and delete operations for red-black trees, we can have the following:

\textbf{4. Red-Black Tree Insertion and Deletion Operations (20 minutes)}
\begin{itemize}
    \item Objective: Apply the insertion and deletion operations.
    \item Breakdown the insertion algorithm, including cases of re-coloring and rotations.
    \item Discuss the deletion process, emphasizing the complexity of fixing the tree afterwards.
    \item Use visual aids to demonstrate both operations and their cases.
    \item Highlight the symmetry between insertion and deletion in terms of maintaining tree properties.
\end{itemize}

\textbf{5. Class Practice and Interactive Problem Solving (10 minutes)}
\begin{itemize}
    \item Divide the students into pairs or small groups and have them work through example problems that require them to insert and delete nodes in a red-black tree.
    \item Use classroom clickers or online polls to choose the correct sequence of operations after each insertion or deletion.
    \item Reconvene and discuss the steps taken and the reasoning behind them, correcting misconceptions and reinforcing correct understanding.
\end{itemize}

This revised outline ensures that students are exposed to both critical operations of red-black trees, understand the challenges associated with each, and can practice applying their knowledge in a controlled, collaborative environment.

\textbf{User:} I will keep your Class Practice and Interactive Problem Solving as an out-of-class activity (most likely a lab session). Please give me the full outline for the class.

\textbf{ChatGPT 4:} Understood. With the Class Practice and Interactive Problem Solving set as a lab session, here is the full revised lecture outline, focusing on both insertion and deletion operations during the class:
Lecture Outline: Red-Black Trees

\begin{enumerate}
    \item Introduction (5 minutes)
    \begin{itemize}
        \item Quick recap on binary search trees.
        \item Introduction to red-black trees and their role in self-balancing data structures.
    \end{itemize}
    
    \item Recap of Red-Black Tree Properties (10 minutes)
    \begin{itemize}
        \item Review the five essential properties of red-black trees.
        \item Illustrate each property with diagrams or animations.
        \item Emphasize the role these properties play in maintaining a balanced tree.
    \end{itemize}
    
    \item Understanding Red-Black Trees (10 minutes)
    \begin{itemize}
        \item Discuss the significance of maintaining red-black tree properties.
        \item Explain the impact on performance for search, insert, and delete operations.
        \item Use visual aids to show how these properties affect tree balance and operation complexity.
    \end{itemize}
    
    \item Red-Black Tree Insertion Operation (10 minutes)
    \begin{itemize}
        \item Step through the insertion algorithm, emphasizing the cases for re-coloring and rotations.
        \item Demonstrate an example insertion, visualizing the adjustments made to maintain properties.
    \end{itemize}
    
    \item Red-Black Tree Deletion Operation (10 minutes)
    \begin{itemize}
        \item Describe the deletion process and the additional complexity compared to insertion.
        \item Walk through a deletion example, showing how the tree compensates for the removed node.
    \end{itemize}
    
    \item Comparison of Insertion and Deletion (5 minutes)
    \begin{itemize}
        \item Summarize and compare the steps and complexity of insertion versus deletion.
        \item Discuss common pitfalls and how to avoid them.
    \end{itemize}
    
    \item Q\&A and Wrap-up (5 minutes)
    \begin{itemize}
        \item Open the floor for questions, encouraging students to think about the operations in depth.
        \item Offer insights into how red-black trees are used in real-world applications.
        \item Assign the interactive problem-solving activity as a lab session.
    \end{itemize}
\end{enumerate}

This outline should cover the essential aspects of red-black trees in the allotted 50 minutes, providing a balanced view of both insertion and deletion operations and setting the stage for a more hands-on exploration in the subsequent lab session.

\textbf{User:} This looks good. Let's work on the specific pieces of the lecture. For the Introduction item, please give me a two-paragraph recap on binary search trees.

\textbf{ChatGPT 4:} Certainly, here's a concise recap on binary search trees for the Introduction part of your lecture:

\textbf{Introduction to Binary Search Trees}
Binary search trees (BSTs) are a fundamental data structure that organizes data in a hierarchical manner for efficient retrieval, insertion, and deletion operations. Each node in a BST contains a unique key, and by definition, the key in any node is larger than the keys in all nodes in its left subtree and smaller than those in its right subtree. This property of binary search trees facilitates efficient searching, akin to a binary search algorithm in a sorted array, where the average time complexity for search operations is O(log n) in a balanced tree. However, this efficiency hinges on the tree's balance; the more unbalanced a tree becomes, with nodes heavily skewed to one side, the closer the search time complexity moves towards O(n), which is no better than a linear search through a list.

The insertion and deletion operations in a BST also follow specific patterns to maintain the binary search property. When inserting a new key, the tree is traversed starting from the root, making a series of decisions to go left or right, depending on whether the key to insert is smaller or larger than the node at each step. Once the correct position is found, the new key is inserted as a leaf node. Deletion, on the other hand, can be more complex, especially when the node to be deleted has two children. In such cases, the node's key is typically replaced with its in-order successor or predecessor, and the successor/predecessor node is then deleted, ensuring the BST properties remain intact. Despite these mechanisms, without additional measures to maintain balance, the performance of BSTs can degrade over time as operations continually add and remove nodes, leading to an unbalanced structure.

\textbf{User:} I like that, but call it a recap of binary search trees.

\textbf{ChatGPT 4:} Absolutely, here's the revised introduction with the specified heading:

\textbf{Recap of Binary Search Trees}
Binary search trees (BSTs) serve as a pivotal data structure for organizing data in a manner that supports efficient data retrieval, insertion, and deletion. At the core of a BST's structure is a key node, which imposes a strict ordering: every node to the left has a smaller key, and every node to the right has a larger key. This intrinsic ordering of BSTs underpins their capability to perform search operations rapidly, akin to a binary search in a sorted array, yielding an average-case time complexity of O(log n) when the tree is balanced. The efficiency of BSTs, however, is contingent on this balanced structure. An unbalanced BST, with a majority of nodes skewed to one side, will deteriorate towards the worst-case time complexity of O(n), analogous to a linear search.

Insertions and deletions within BSTs are operations bound by the need to maintain the binary search property. Insertion begins at the root, traversing the tree and choosing a direction—left for a smaller key and right for a larger key—until reaching an appropriate leaf position for the new key. Conversely, deletion is inherently more complex, especially when the target node has two children. The common strategy involves substituting the key of the node to be deleted with its in-order successor or predecessor and then recursively deleting that node. These procedures ensure the BST's structural integrity is preserved. Nonetheless, the inherent challenge lies in maintaining the balance of the tree through successive insertions and deletions, as an unbalanced tree can significantly hinder the BST's performance capabilities.

\textbf{User:} Let's keep working on the introduction part of the lecture. Please give me an Introduction to red-black trees and their role in self-balancing data structures.

\textbf{ChatGPT 4:} \textbf{Introduction to Red-Black Trees}
Red-black trees are an ingenious type of self-balancing binary search tree, a sophisticated data structure that overcomes the fundamental issue of maintaining balance after insertion and deletion operations. Each node in a red-black tree stores an extra bit for denoting the color of the node, either red or black, which is instrumental in enforcing a balanced tree structure. This color attribute, along with specific rules pertaining to node colors and black heights, ensures that the tree remains approximately balanced at all times. As a result, red-black trees guarantee that the tree height remains logarithmic in relation to the number of nodes, thus ensuring that search, insert, and delete operations can be performed in O(log n) time, even in the worst-case scenario.

The real-world significance of red-black trees is profound, particularly in computer science applications that require rapid data look-up and modification, such as database indexes and associative arrays. The design of red-black trees addresses the imbalance that can occur in simple binary search trees, where the depth can become linear in the number of elements, leading to inefficient operations. By automatically rearranging themselves during insertions and deletions to preserve balance, red-black trees provide a reliable foundation for dynamic data storage and retrieval systems. Their ability to self-balance with minimal overhead makes them an invaluable tool in the repertoire of data structures, ensuring that they remain a critical component in the efficient implementation of various abstract data types.

\textbf{User:} Good. The second part of the lecture is about formal definitions and properties of red-black trees. Let's start with the five essential properties of red-black trees. Please describe them formally. Remember, you are a professor of computer science from a top 10 engineering university.

\textbf{ChatGPT 4:} \textbf{Formal Definitions and Properties of Red-Black Trees}
In red-black trees, the structure is maintained by a set of rules—referred to as properties—that must be preserved following all insertion and deletion operations. Here are the five essential properties that define a red-black tree:

\begin{enumerate}
    \item Property One: Node Coloring Every node is colored either red or black. This binary coloring is crucial for maintaining a balanced tree and is a critical part of the red-black tree's self-balancing properties.
    \item Property Two: Root Blackness The root of the tree is always black. This property ensures that the path from the root to the leaves has a consistent number of black nodes, which is fundamental to maintaining balance.
    \item Property Three: Red Node Children All red nodes must have black children (also known as the "no two consecutive red nodes" rule). This property prevents the formation of red "runs" in the tree, which would indicate an imbalance.
    \item Property Four: Uniform Black Height Every path from a node to its descendant NULL nodes must have the same number of black nodes. This uniform number of black nodes is known as the "black height," and its consistency is key to maintaining balance across the tree.
    \item Property Five: New Node Insertions All new nodes are inserted as red nodes. While this might momentarily disrupt the tree's balance, it simplifies the rebalancing process that follows the insertion.
\end{enumerate}

These properties collectively ensure that the longest path from the root to the farthest leaf is no more than twice as long as the shortest path, which implies that the tree remains approximately balanced. This balance is what allows the red-black tree to ensure a worst-case time complexity of O(log n) for search, insertion, and deletion operations, thus maintaining efficient performance.

\textbf{User:} Your property five is not an actual property of red-black trees but an important consideration for the insertion algorithm. The property you are missing says that null nodes are black. Please reformulate the properties.

\textbf{ChatGPT 4:} Apologies for the oversight. Let's correct the formal definition and properties of red-black trees:

\textbf{Formal Definitions and Properties of Red-Black Trees}
A red-black tree is a binary search tree with an additional bit of storage per node: its color, which can be either red or black. The colors of the nodes are used to ensure the tree remains balanced during insertions and deletions. Below are the five essential properties that every red-black tree must satisfy:

\begin{enumerate}
    \item Property One: Node Coloring Each node is either red or black. This property is fundamental to the operation of red-black trees, as the color assigned to each node is pivotal to enforcing balance throughout the structure.
    \item Property Two: Root Blackness The root of the tree is always black. By convention, the root's color is black to maintain consistent black depth across the tree.
    \item Property Three: Red Node Constraints Red nodes cannot have red children, a rule often stated as "a red node cannot have a red parent". This ensures that no path will have two consecutive red nodes, thus preventing significant imbalances in the tree's depth.
    \item Property Four: Black Height Consistency Every path from a node to any of its descendant NIL nodes goes through the same number of black nodes. This number of black nodes is called the black height. This property guarantees that the tree remains balanced, with the path from the root to the farthest leaf not more than twice as long as the shortest path.
    \item Property Five: NIL Nodes as Black All leaves (NIL nodes) are black. In a red-black tree, the leaves are typically represented as NULL, which are considered black nodes. This convention simplifies the rules for red nodes and black height since NULL nodes end up contributing to the black height of the tree.
\end{enumerate}

The enforcement of these properties after every insertion and deletion is what allows the red-black tree to self-balance. It is the interaction of these properties that ensures the operations on red-black trees can be done in $O(\log{n})$ time, which is critical for maintaining the high performance of the data structures that utilize them.

\textbf{User:} Thanks. To guarantee the formality of the definitions, please give me the mathematical proof for the claim that, by maintaining the black height of the tree, its respective height is $O(\log{n})$.

\textbf{ChatGPT 4:} To prove that the height of a red-black tree is $O(\log{n})$ by maintaining the black height, we consider the relationship between the number of nodes in the tree and the "black height" - the number of black nodes on any path from the root to a leaf.

\textbf{Definitions:}
\begin{itemize}
    \item Let $n$ be the number of nodes in the red-black tree.
    \item Let $h$ be the height of the tree (the longest path from the root to a leaf).
    \item Let $bh$ be the black height of the tree (the number of black nodes on any path from the root to a leaf).
\end{itemize}

\textbf{Properties Used in the Proof:}
\begin{enumerate}
    \item A red node cannot have a red child (Property 3).
    \item All paths from the root to the leaves contain the same number of black nodes (Property 4).
\end{enumerate}

\textbf{Lemma:} A red-black tree with black height $bh$ has at least $2^{bh}-1$ internal nodes.

\textbf{Proof of Lemma:}
\begin{itemize}
    \item The shortest possible path from the root to a leaf has all black nodes (by Property 4). There are bhbh black nodes on this path.
    \item The longest possible path alternates between red and black nodes. Because of Property 3, the path cannot have two consecutive red nodes, so the path has at most 
    $2^{bh}-1$ nodes since it could start with a red node (the root is always black, but the child can be red) and alternate from there.
    \item The smallest number of nodes in a perfectly balanced binary tree of height $h$ (where $h$ is the number of edges on the longest path from the root to a leaf) is $2^{h}-1$ (this is a property of a complete binary tree).
\end{itemize}

\textbf{Proof of Height Complexity:}
\begin{itemize}
    \item Because all paths have the same number of black nodes, the shortest path has $bh$ nodes and the longest path can have at most twice as many, $2^{bh}$.
    \item So, in the worst case, the height $h$ of the tree is $2^{bh}$.
    \item From the lemma, a tree with black height $bh$ has at least $2^{bh}-1$ internal nodes. Therefore, $n\geq2^{bh}-1$.
    \item Solving for $bh$, we get $bh\leq\log_{2}{n+1}$.
    \item Therefore, $h \leq 2^{bh} \leq 2\log_{2}{n+1}$, which simplifies to $h=O(\log{n})$.
\end{itemize}

Thus, the height of a red-black tree is logarithmically bound by the number of nodes in the tree, proving that operations such as search, insert, and delete can be performed in $O(\log{n})$ time.

\textbf{User:} This is great but also too much for the 20 minutes of the class allotted towards the last two sections of the class. Please remind me of the outline for these two sections and give me a shorter script for them.

\textbf{ChatGPT 4:} Certainly, let's revisit the outline for the last two sections of your class on red-black trees and provide a succinct script for each part.

\textbf{Lecture Outline:}
\begin{enumerate}
    \item Red-Black Tree Insertion Operation (10 minutes)
    \begin{itemize}
        \item Discuss the insertion process.
        \item Walk through the steps of fixing violations after insertion.
    \end{itemize}
    
    \item Red-Black Tree Deletion Operation (10 minutes)
    \begin{itemize}
        \item Explain the deletion process.
        \item Outline the steps of rebalancing the tree after deletion.
    \end{itemize}
\end{enumerate}

\textbf{Shorter Script for Section 4: Red-Black Tree Insertion Operation}
"In the next ten minutes, we'll unpack the insertion operation for red-black trees. Insertions begin as they would in a standard binary search tree, with the new node being initially inserted as a red node to maintain balance more easily. However, this can cause violations of our red-black properties, specifically the potential for two consecutive red nodes. To resolve these issues, we perform a series of color changes and tree rotations. We categorize these fixes into three main cases:

\begin{itemize}
    \item Case 1 deals with a red uncle node and is resolved by recoloring.
    \item Case 2 and Case 3 handle a black uncle node, requiring rotations — a single rotation for Case 2 and a double rotation for Case 3. Each case is a step towards re-establishing our red-black properties, ensuring the tree remains balanced. Let's quickly walk through an example..."
\end{itemize}

\textbf{Shorter Script for Section 5: Red-Black Tree Deletion Operation}
"Moving on to deletion, which we have the remaining ten minutes to cover. Deletion in a red-black tree is more intricate due to the need to replace the deleted node and then fix any potential property violations. When a node is removed, we may be left with a 'double black' issue, where a black node has been removed from a black parent, causing an imbalance. To correct this, we have several cases to consider, similar to insertion:

\begin{itemize}
    \item If a sibling node is red, we perform a rotation to move the problem down the tree.
    \item If the sibling and its children are black, we recolor and move up the tree to address the double black at the parent.
    \item If the sibling has a red child, we perform rotations and recolorings to balance the tree. Let's illustrate this with a simple deletion example, focusing on how these operations restore our crucial properties."    
\end{itemize}

By condensing the explanations and focusing on the high-level process rather than delving into all the intricate details, you'll be able to cover both insertion and deletion in the allotted 20 minutes, leaving room for examples and student engagement.

\textbf{User:} Please be careful. The sections we were working on were the recap of Red-Black tree properties and the Understanding Red-Black trees section. Please remind me and yourself of the approved outline for the lecture (don't forget, you're acting like a professor of computer science from a top-10 engineering university).

\textbf{ChatGPT 4:} I apologize for the confusion. Let's revisit the correct sections of your lecture outline on red-black trees and provide a concise script for each.

\textbf{Approved Lecture Outline:}
\begin{enumerate}
    \item Introduction (5 minutes)
    \begin{itemize}
        \item Recap of binary search trees.
    \end{itemize}
    
    \item Recap of Red-Black Tree Properties (10 minutes)
    \begin{itemize}
        \item Formal definitions and descriptions of the five essential properties of red-black trees.
    \end{itemize}
    
    \item Understanding Red-Black Trees (10 minutes)
    \begin{itemize}
        \item Discussion on the significance of red-black tree properties and their impact on the tree's operations and performance. 
    \end{itemize}
\end{enumerate}

\textbf{Shorter Script for Section 2: Recap of Red-Black Tree Properties}
"In the next ten minutes, we'll revisit the formal properties that define red-black trees. These properties are pivotal to ensuring the self-balancing nature of these trees. They are:

\begin{enumerate}
    \item Node Color: Every node is either red or black, creating the foundation for our balancing rules.
    \item Root Property: The root of the tree is always black, which anchors our black path-lengths.
    \item Red Node Property: If a node is red, then both its children must be black. This prevents consecutive red nodes.
    \item Black Height Property: Every path from a node to its descendant leaves has the same number of black nodes, which we refer to as the black height.
    \item Leaf Nodes: All leaves (NIL nodes) are black, simplifying the enforcement of the tree properties.
\end{enumerate}

Remember, these rules work in concert to maintain a balanced tree, ensuring that the longest path from the root to a leaf is no more than twice the length of the shortest path."

\textbf{Shorter Script for Section 3: Understanding Red-Black Trees}

"Now, let's delve into the significance of these properties for the next ten minutes. The essence of a red-black tree's efficiency lies in its balanced structure. The stipulation that each path from root to leaf has the same black height prevents the formation of disproportionately long branches. This uniformity guarantees that the worst-case height of the tree remains logarithmic relative to the number of nodes.

When we insert or delete nodes, the tree's self-balancing properties necessitate a series of color changes and rotations to restore balance. But why do we care about balance? It's simple: balance translates to efficiency. A balanced tree ensures that the time complexity for search, insertion, and deletion remains at $O(\log{n})$, which is crucial for high-performance computing tasks. Keep in mind, a red-black tree doesn't aim to be perfectly balanced; rather, it aims to be balanced enough to guarantee this efficiency, which is a beautiful instance of practical optimization."

This script covers the fundamental properties and understanding of red-black trees concisely, while respecting the allotted time for each section.

\textbf{User:} To make sure that we are complete with the first parts of the lecture, please give me a short outline of the proof about the height or red-black trees to be $O(\log{n})$.

\textbf{ChatGPT 4:} Certainly, a concise outline of the proof that the height of red-black trees is $O(\log{n})$ would include the following points:

\textbf{Outline of Proof:}
\begin{enumerate}
    \item 	Black Height and Complete Trees:
    \begin{itemize}
    \item Define black height (the number of black nodes from the root to a leaf, not counting the leaf node itself).
    \item Relate black height to a complete binary tree, which is a perfect tree where all levels are fully filled.
    \item State that the minimum number of nodes in a complete binary tree of black height $bh$ is $2^{bh}-1$.
    \end{itemize}
    
    \item 	Red-Black Tree Properties:
    \begin{itemize}
    \item Reiterate the red-black tree properties relevant to the proof, particularly the property that all paths from the root to the leaves have the same number of black nodes.
    \end{itemize}
    
    \item 	Shortest and Longest Paths:
    \begin{itemize}
    \item Explain that the shortest possible path from the root to any leaf in a red-black tree has $bh$ nodes, all of which are black.
    \item Argue that the longest path can have at most double the number of black nodes since it can alternate between red and black nodes (but cannot have two consecutive red nodes).
    \end{itemize}
    
    \item 	Counting Nodes:
    \begin{itemize}
    \item Use the properties to show that the shortest path in a red-black tree is at least half the length of the longest path.
    \item Deduce that the number of nodes $n$ in the tree satisfies $N \geq 2^{bh}-1$.
    \end{itemize}
    
    \item 	Logarithmic Height:
    \begin{itemize}
    \item Take logarithms to show $bh \le log2(n+1)$.
    \item Since the height h is less than twice the black height, conclude $h<2bh \le 2log(n+1)$.
    \item Simplify to show that $h=O(log(n))$.
    \end{itemize}
\end{enumerate}

This outline provides a structured approach to prove that the height of a red-black tree is logarithmic with respect to the number of nodes, keeping the explanation high-level and suitable for a brief lecture segment.

\textbf{User:} That is good. Thanks! Let's work on the part of the lecture dedicated to the insert and delete operations. For the insert operation, please give me its main motivation and pseudocode.

\textbf{ChatGPT 4:} \textbf{Insert Operation in Red-Black Trees}
Motivation for Insertion Operation: The primary motivation for the insertion operation in red-black trees is to add a new node while preserving the red-black properties. Since direct insertion could violate these properties, particularly the red node property and black height uniformity, an insertion algorithm must include a re-balancing procedure. This re-balancing is crucial to maintain the tree's balanced state, which in turn ensures that the tree's operations remain efficient, with a time complexity of $O(log(n))$.
Pseudocode for Insertion Operation:

\begin{verbatim}
RB-INSERT(T, z)
    // T is the red-black tree, and z is the node to be inserted
    y = NIL
    x = T.root
    while x \neq NIL
        y = x
        if z.key < x.key
            x = x.left
        else
            x = x.right
    z.parent = y
    if y = NIL
        T.root = z // Tree T was empty
    else if z.key < y.key
        y.left = z
    else
        y.right = z
    z.left = NIL
    z.right = NIL
    z.color = RED // Insert node as red
    RB-INSERT-FIXUP(T, z) // Rebalance the tree
\end{verbatim}

\begin{verbatim}
RB-INSERT-FIXUP(T, z)
    while z.parent.color == RED
        if z.parent == z.parent.parent.left
            y = z.parent.parent.right // y is z's uncle
            if y.color == RED
                // Case 1: z's uncle y is red
                z.parent.color = BLACK
                y.color = BLACK
                z.parent.parent.color = RED
                z = z.parent.parent
            else
                if z == z.parent.right
                    // Case 2: z is a right child
                    z = z.parent
                    LEFT-ROTATE(T, z)
                // Case 3: z is a left child
                z.parent.color = BLACK
                z.parent.parent.color = RED
                RIGHT-ROTATE(T, z.parent.parent)
        else
            // Same as above with "right" and "left" exchanged
    T.root.color = BLACK // Ensure root is black
\end{verbatim}

The RB-INSERT-FIXUP function corrects the red-black properties that might have been violated during insertion. This involves a series of checks and adjustments, including potential color changes and rotations (left or right), which are classified into several cases:

\begin{itemize}
    \item Case 1: Red uncle, which is resolved by recoloring.
    \item Case 2: Black uncle and the new node is a right child, requiring a left rotation.
    \item Case 3: Black uncle and the new node is a left child, requiring a right rotation.
\end{itemize}

The exact operations performed depend on the structure of the tree at the time of insertion. This procedure ensures that the red-black properties are preserved, maintaining the tree's balanced state.

\textbf{User:} Please do the same for the delete operation.

\textbf{ChatGPT 4:} \textbf{Delete Operation in Red-Black Trees}
Motivation for Deletion Operation: The deletion operation in red-black trees is crucial for removing nodes while preserving the tree's balanced state. A straightforward removal could disrupt the red-black properties, especially if a black node is removed, potentially unbalancing the tree. This unbalance could degrade the performance of the tree's operations. Hence, the deletion algorithm includes a meticulous re-balancing process called 'fix-up' to adjust colors and perform rotations to uphold the tree's properties and efficiency.

\textbf{Pseudocode for Deletion Operation:}

\begin{verbatim}
RB-DELETE(T, z)
    // T is the red-black tree, and z is the node to be deleted
    y = z
    y_original_color = y.color
    if z.left == NIL
        x = z.right
        RB-TRANSPLANT(T, z, z.right)
    else if z.right == NIL
        x = z.left
        RB-TRANSPLANT(T, z, z.left)
    else
        y = TREE-MINIMUM(z.right)
        y_original_color = y.color
        x = y.right
        if y.parent == z
            x.parent = y
        else
            RB-TRANSPLANT(T, y, y.right)
            y.right = z.right
            y.right.parent = y
        RB-TRANSPLANT(T, z, y)
        y.left = z.left
        y.left.parent = y
        y.color = z.color
    if y_original_color == BLACK
        RB-DELETE-FIXUP(T, x)
\end{verbatim}

\begin{verbatim}
RB-DELETE-FIXUP(T, x)
    while x \neq T.root and x.color == BLACK
        if x == x.parent.left
            w = x.parent.right
            if w.color == RED
                w.color = BLACK
                x.parent.color = RED
                LEFT-ROTATE(T, x.parent)
                w = x.parent.right
            if w.left.color == BLACK and w.right.color == BLACK
                w.color = RED
                x = x.parent
            else
                if w.right.color == BLACK
                    w.left.color = BLACK
                    w.color = RED
                    RIGHT-ROTATE(T, w)
                    w = x.parent.right
                w.color = x.parent.color
                x.parent.color = BLACK
                w.right.color = BLACK
                LEFT-ROTATE(T, x.parent)
                x = T.root
        else
            // Same as then clause with "right" and "left" exchanged
    x.color = BLACK    
\end{verbatim}

The RB-DELETE-FIXUP function is designed to restore the red-black properties that might have been violated during deletion. This involves several cases that are addressed depending on the colors of the sibling and the sibling's children, which may require recoloring and rotations. The goal is to redistribute or adjust the black heights so that the tree remains balanced after the deletion:

\begin{itemize}
    \item If the sibling is red, a rotation is performed to set up further adjustments.
    \item If the sibling and its children are black, recoloring occurs and the problem ascends the tree.
    \item If the sibling has a red child, rotations and recoloring are performed to fix the imbalance.
\end{itemize}

This fix-up procedure ensures the red-black properties are intact after deletion, maintaining the tree's balanced state and operation efficiency.

\textbf{User:} This is good. Please remind me of the class outline again.

\textbf{ChatGPT 4:} Certainly! Here's the class outline we've established for your lecture on red-black trees:

\textbf{Lecture Outline: Red-Black Trees}

\begin{enumerate}
    \item Introduction (5 minutes)
    \begin{itemize}
        \item A recap on binary search trees.
    \end{itemize}
    
    \item Recap of Red-Black Tree Properties (10 minutes)
    \begin{itemize}
        \item Formal definitions and descriptions of the five essential properties of red-black trees.
    \end{itemize}
    
    \item Understanding Red-Black Trees (10 minutes)
    \begin{itemize}
        \item Discussion on the significance of red-black tree properties and their impact on the tree's operations and performance.
    \end{itemize}
    
    \item Red-Black Tree Insertion Operation (10 minutes)
    \begin{itemize}
        \item Main motivation for the insertion operation.
        \item Explanation of the insertion process and pseudocode overview.
    \end{itemize}
    
    \item Red-Black Tree Deletion Operation (10 minutes)
    \begin{itemize}
        \item Main motivation for the deletion operation.
        \item Explanation of the deletion process and pseudocode overview.
    \end{itemize}
    
    \item Q\&A and Wrap-up (5 minutes)
    \begin{itemize}
        \item Open the floor for questions and engage with the students to clarify any doubts.
        \item Summarize the key points covered in the lecture.
        \item Mention of out-of-class activities related to red-black trees, such as lab sessions or problem sets.
    \end{itemize}
    
\end{enumerate}

This outline is designed to cover the fundamentals and complexities of red-black trees within a single lecture session, providing a balance between theoretical understanding and practical application.

\textbf{User:} Awesome. Let's wrap it up and give me the summary of the lecture.

\textbf{ChatGPT 4:} \textbf{Lecture Summary: Red-Black Trees}
Today, we revisited the fundamental aspects of binary search trees, highlighting their efficiency and the challenges faced when they become unbalanced. We then introduced red-black trees, a self-balancing variant of binary search trees, which mitigate these challenges through a set of well-defined properties.

We delved into the five essential properties of red-black trees:
\begin{enumerate}
    \item Each node is colored either red or black.
    \item The root is always black.
    \item Red nodes cannot have red children.
    \item All paths from a node to its descendant leaves contain the same number of black nodes.
    \item All leaves (NIL nodes) are black.
\end{enumerate}

Understanding these properties is crucial as they ensure the tree remains balanced after every insertion and deletion, thereby maintaining the operations' time complexity at O(log(n)).

We covered the insertion operation, which adds a node to the tree and then enforces the red-black properties through a series of rotations and color changes if necessary. We also discussed the more complex deletion operation, which removes a node and then performs a fix-up process to restore the tree's balanced state.

The motivation behind both operations is to maintain the high efficiency of the tree’s search, insert, and delete operations by preserving its structured balance. The pseudocode provided offers a blueprint for implementing these operations, reinforcing the algorithmic precision required to maintain the integrity of the red-black trees.

In conclusion, red-black trees are a robust data structure, ideal for applications that require efficient data storage and retrieval. The balancing of the tree is fundamental to its performance, and understanding how to maintain this balance through insertion and deletion operations is key to utilizing red-black trees effectively.

Remember, today’s lecture content is foundational for your upcoming lab session, where you will apply these concepts in practice. Please review the pseudocode and properties discussed as preparation for the hands-on activities. Thank you for your attention, and I look forward to seeing how you all tackle the intricacies of red-black trees in our lab.

\textbf{User:} Please compile the learning objectives, outline, and content into a single large lecture document.

\textbf{ChatGPT 4:} \textbf{Lecture Document: Understanding Red-Black Trees}
Learning Objectives: By the end of this lecture, students should be able to:

\begin{enumerate}
    \item Recall and describe the properties that define a red-black tree.
    \item Explain the significance of each red-black tree property in maintaining a balanced tree and how it affects the tree's performance for operations such as insertion and deletion.
    \item Implement the insertion operation in a red-black tree, ensuring that the tree maintains its defining properties after adding a new node.
    \item Outline and comprehend the deletion process in a red-black tree and the steps taken to fix any violations of the tree's properties.
\end{enumerate}

\textbf{Lecture Outline:}
\begin{enumerate}
    \item Introduction (5 minutes)
    \begin{itemize}
        \item Recap of binary search trees.
    \end{itemize}
    
    \item Recap of Red-Black Tree Properties (10 minutes)
    \begin{itemize}
        \item Formal definitions and descriptions of the five essential properties of red-black trees.
    \end{itemize}
    
    \item Understanding Red-Black Trees (10 minutes)
    \begin{itemize}
        \item Discussion on the significance of red-black tree properties and their impact on the tree's operations and performance.
    \end{itemize}
    
    \item Red-Black Tree Insertion Operation (10 minutes)
    \begin{itemize}
        \item Main motivation for the insertion operation.
        \item Explanation of the insertion process and pseudocode overview.
    \end{itemize}
    
    \item Red-Black Tree Deletion Operation (10 minutes)
    \begin{itemize}
        \item Main motivation for the deletion operation.
        \item Explanation of the deletion process and pseudocode overview.
    \end{itemize}
    
    \item Q\&A and Wrap-up (5 minutes)
    \begin{itemize}
        \item Open the floor for questions and engage with the students to clarify any doubts.
        \item Summarize the key points covered in the lecture.
        \item Mention of out-of-class activities related to red-black trees, such as lab sessions or problem sets.
    \end{itemize}
\end{enumerate}

\textbf{Content:}
\textbf{1. Introduction to Binary Search Trees} Binary search trees (BSTs) are a cornerstone of efficient data organization, allowing for quick data retrieval, insertion, and deletion. BSTs maintain a specific order, where each node's left descendants are less than the node and right descendants are greater. While BSTs are efficient with a balanced tree, the cost of operations can degrade to O(n) if the tree becomes unbalanced.

\textbf{2. Recap of Red-Black Tree Properties} Red-black trees enhance BSTs by ensuring the tree remains balanced with each operation through these properties:
\begin{itemize}
    \item Nodes are either red or black.
    \item The root is always black.
    \item Red nodes have black children.
    \item Every path from a node to its descendant NIL nodes has the same number of black nodes.
    \item All NIL nodes are black.
\end{itemize}

\textbf{3. Understanding Red-Black Trees} The properties of red-black trees work together to maintain a balanced height, ensuring that operations remain efficient at O(log n) complexity. This balance is key to the performance and reliability of red-black trees in various applications.

\textbf{4. Red-Black Tree Insertion Operation} The insertion operation in red-black trees starts by adding the node as a red node to maintain balance and then making necessary adjustments:
\begin{itemize}
    \item If a violation occurs, we perform rotations and recolorings to fix the tree.
    \item The goal is to maintain the essential properties of the red-black tree through this process.
\end{itemize}

\textbf{Insertion Pseudocode:}
\begin{verbatim}
RB-INSERT(T, z)
    ...
    RB-INSERT-FIXUP(T, z)
    ...    
\end{verbatim}

\textbf{5. Red-Black Tree Deletion Operation} Deletion involves removing a node and then ensuring the tree remains balanced by addressing any property violations through a series of rotations and color changes.

\textbf{Deletion Pseudocode:}
\begin{verbatim}
RB-DELETE(T, z)
    ...
    RB-DELETE-FIXUP(T, x)
    ...    
\end{verbatim}

\textbf{6. Q\&A and Wrap-up} We concluded the lecture with a discussion and clarification session, encouraging students to think critically about the mechanisms of red-black trees and preparing them for practical applications in the upcoming lab.

This document encapsulates the goals, structure, and detailed content of the lecture on red-black trees. It should serve as a comprehensive guide for both teaching and review.

\section{Bing Copilot} \label{sec_bingco}
This is a nearly identical conversation (from the user's side) with Bing Copilot as we had with ChatGPT 4.0.

Conversation took place in November of 2023.

\textbf{User:} Act as a professor of computer science from a top 10 engineering university. You are teaching a course about data structures and algorithms to Computer Science and Data Science majors (students take this course in their sophomore year). So far, you have covered primitive data structures, asymptotic runtime analysis, and dynamic data structures. The last class was about binary search trees, emphasizing the problems we could get with them. Please give me five learning objectives for a class discussing red-black trees. Use Bloom’s revised taxonomy of verbs for the objectives.

\textbf{Bing Copilot:} Sure, here are five learning objectives for a class discussing red-black trees using Bloom's revised taxonomy of verbs:
\begin{enumerate}
    \item Understand the concept of red-black trees: Students should be able to explain what red-black trees are, their properties, and why they are used in computer science.
    \item Analyze the structure of red-black trees: Students should be able to dissect a given red-black tree, identify its components, and explain the role of each component.
    \item Apply the operations of red-black trees: Students should be able to perform basic operations on red-black trees such as insertion, deletion, and searching.
    \item Evaluate the performance of red-black trees: Students should be able to assess the time complexity of operations on red-black trees and compare it with other data structures.
    \item Create red-black trees: Given a set of data, students should be able to construct a red-black tree.
\end{enumerate}
These objectives aim to provide a comprehensive understanding of red-black trees, their structure, operations, performance, and practical applications. They are designed to promote higher-order thinking skills as per Bloom's revised taxonomy.

\textbf{User:} I like your first three learning objectives! With these in mind, please propose a lecture outline for a 50-minute class.

\textbf{Bing Copilot:} Sure, here's a proposed lecture outline for a 50-minute class on red-black trees:
\begin{enumerate}
    \item Introduction (5 minutes): Start the class with a brief recap of binary search trees and their limitations that were discussed in the last class. Introduce the concept of red-black trees as a solution to these problems.
    \item Understanding Red-Black Trees (10 minutes): Explain the properties of red-black trees and how they ensure balance in the tree. Discuss the color property of nodes and why it is essential.
    \item Break (5 minutes): A short break for students to digest the information.
    \item Analyzing Red-Black Trees (10 minutes): Use diagrams and examples to analyze the structure of red-black trees. Discuss the significance of black depth.
    \item Applying Operations on Red-Black Trees (15 minutes):
    \begin{itemize}
        \item Insertion (7 minutes): Explain and demonstrate how to insert a node into a red-black tree. Discuss how the tree maintains its properties after insertion.
        \item Deletion (8 minutes): Explain and demonstrate how to delete a node from a red-black tree. Discuss how the tree maintains its properties after deletion.
    \end{itemize}
    \item Summary and Q\&A (5 minutes): Summarize the key points of the lecture and open the floor for questions.
\end{enumerate}
This outline ensures that the class is interactive and gives students the opportunity to understand, analyze, and apply their knowledge of red-black trees. It also includes a break to help maintain the students' concentration and engagement levels. The Q\&A session at the end allows for clarification of doubts and deeper understanding.

\textbf{User:} I do not think we need to give the students a break (they are used to dense lectures). Could you please rewrite the outline?

\textbf{Bing Copilot:} Sure, here's a revised lecture outline for a 50-minute class on red-black trees:
\begin{enumerate}
    \item Introduction (5 minutes): Start the class with a brief recap of binary search trees and their limitations that were discussed in the last class. Introduce the concept of red-black trees as a solution to these problems.
    \item Understanding Red-Black Trees (10 minutes): Explain the properties of red-black trees and how they ensure balance in the tree. Discuss the color property of nodes and why it is essential.
    \item Analyzing Red-Black Trees (10 minutes): Use diagrams and examples to analyze the structure of red-black trees. Discuss the significance of black depth.
    \item Applying Operations on Red-Black Trees (15 minutes):
    \begin{itemize}
        \item Insertion (7 minutes): Explain and demonstrate how to insert a node into a red-black tree. Discuss how the tree maintains its properties after insertion.
        \item Deletion (8 minutes): Explain and demonstrate how to delete a node from a red-black tree. Discuss how the tree maintains its properties after deletion.
    \end{itemize}
    \item Summary and Q\&A (10 minutes): Summarize the key points of the lecture and open the floor for questions.
\end{enumerate}
This outline ensures that the class is interactive and gives students the opportunity to understand, analyze, and apply their knowledge of red-black trees. The Q\&A session at the end allows for clarification of doubts and deeper understanding.

\textbf{User:} This looks good. Let's work on the specific pieces of the lecture. For the Introduction item, please give me a two-paragraph recap on binary search trees.

\textbf{Bing Copilot:} Sure, here's a two-paragraph recap on binary search trees:

A binary search tree (BST) is a node-based binary tree data structure that has the following properties: The left subtree of a node contains only nodes with keys lesser than the node’s key. The right subtree of a node contains only nodes with keys greater than the node’s key. The left and right subtree each must also be a binary search tree$^2$. Binary search trees are used to store data in a way that allows for efficient searching, insertion, and deletion of elements. They are commonly used in computer science and are the foundation for many other data structures.

Binary search trees are useful because they allow for fast searching of elements. The search operation in a binary search tree is performed by comparing the search key with the key of the root node. If the search key is less than the key of the root node, the search continues in the left subtree. If the search key is greater than the key of the root node, the search continues in the right subtree. This process is repeated until the search key is found or the tree is exhausted$^2$. Binary search trees are also used to implement other data structures such as sets, multisets, and associative arrays.

Source(s)
\begin{enumerate}
    \item \href{https://en.wikipedia.org/wiki/Binary_search_tree}{Binary search tree - Wikipedia}
    \item \href{https://www.geeksforgeeks.org/binary-search-tree-data-structure/}{Binary Search Tree - GeeksforGeeks}
    \item \href{https://www.baeldung.com/cs/binary-search-trees}{A Quick Guide to Binary Search Trees - Baeldung}
    \item \href{https://algs4.cs.princeton.edu/32bst/}{Binary Search Trees - Princeton University}
    \item \href{https://www.geeksforgeeks.org/introduction-to-binary-search-tree-data-structure-and-algorithm-tutorials/}{Introduction to Binary Search Tree - GeeksforGeeks}
    \item \href{https://www.programiz.com/dsa/binary-search-tree}{Binary Search Tree - Programiz}
    \item \href{http://cslibrary.stanford.edu/110/BinaryTrees.html}{Binary Trees - Stanford University}
    \item \href{https://en.wikipedia.org/wiki/Binary_search_tree}{en.wikipedia.org}
\end{enumerate}

\textbf{User:} Let's keep working on the introduction part of the lecture. Please give me an Introduction to red-black trees and their role in self-balancing data structures.

\textbf{Bing Copilot:} Red-black trees are a type of self-balancing binary search tree that use a specific set of rules to ensure that the tree is always balanced. This balance guarantees that the time complexity for operations such as insertion, deletion, and searching is always O(log n), regardless of the initial shape of the tree$^1$.

Red-black trees are similar to binary search trees, but with an additional color property for each node. Each node in a red-black tree is either red or black. The color of a node is used to ensure that the tree remains balanced. The rules for coloring nodes are as follows:
\begin{itemize}
    \item The root node must be black.
    \item Every leaf node (NULL node) must be black.
    \item If a node is red, its children must be black.
    \item Every path from a node to its descendant leaf nodes must contain the same number of black nodes.
\end{itemize}
These rules ensure that the longest path from the root to a leaf node is no more than twice as long as the shortest path$^1$. This balance guarantees that the time complexity for operations such as insertion, deletion, and searching is always O(log n), regardless of the initial shape of the tree$^1$. Red-black trees are used in many applications, including memory allocation algorithms, compilers, and databases$^2$.

Source(s)
\begin{enumerate}
    \item \href{https://www.geeksforgeeks.org/introduction-to-red-black-tree/}{Introduction to Red-Black Tree - GeeksforGeeks}
    \item \href{https://bing.com/search?q=red-black+trees+introduction}{Red–black tree - Wikipedia}
    \item \href{https://en.wikipedia.org/wiki/Red%E2%80%93black_tree}{Red–black tree - Wikipedia}
    \begin{itemize}
        \item \textit{Author note: this is copied verbatim, including the urls.}
    \end{itemize}
    \item \href{https://redirect.cs.umbc.edu/courses/undergraduate/341/fall13/section3/lectures/10-Red-Black-Trees.pdf}{Red-Black Trees - Department of Computer Science and Electrical Engineering}
    \item \href{}{Applications of Red-Black Trees | Baeldung on Computer Science}
\end{enumerate}

\textbf{User:} Thanks. To guarantee the formality of the definitions, please give me the mathematical proof for the claim that, by maintaining the black height of the tree, its respective height is $O(log(n))$. Remember we are in a lecture with a time limit.

\textbf{Bing Copilot:} The height of a red-black tree is guaranteed to be $O(log n)$ by maintaining the black height of the tree. The black height of a node is defined as the number of black nodes on any simple path from that node to a leaf node. The black height of all leaf nodes is zero. The black height of a red-black tree is the black height of its root node.

To prove that the height of a red-black tree is $O(log n)$, we need to show that the maximum height of the tree is proportional to the logarithm of the number of nodes in the tree. We can do this by showing that the number of nodes in a red-black tree of height h is at most $2^{2h+1}-1$.

We will prove this by induction. For the base case, consider a red-black tree of height 0. This tree has only one node, which is the root node. Therefore, the number of nodes in the tree is 1, which is equal to $2^{2*0+1}-1$.

Now assume that the claim is true for all red-black trees of height less than h. Consider a red-black tree of height h. By the definition of a red-black tree, the root node must be black. Therefore, the two children of the root node must be red or black. If they are both black, then the black height of the tree is h-1, and the number of nodes in the left and right subtrees is at most $2^{2(h-1)+1}-1$ by the induction hypothesis. Therefore, the total number of nodes in the tree is at most $2^{2(h-1)+1}-1 + 1 + 2^{2(h-1)+1}-1 = 2^{2h+1}-1$.

If the two children of the root node are both red, then the black height of the tree is h, and the number of nodes in the left and right subtrees is at most $2^{2(h-1)}-1$ by the induction hypothesis. Therefore, the total number of nodes in the tree is at most $2^{2(h-1)}-1 + 1 + 2^{2(h-1)}-1 = 2^{2h}-1$.

In either case, the number of nodes in the tree is at most $2^{2h+1}-1$. Therefore, the height of the tree is at most $log(2^{2h+1}-1) = O(log n)$.

This completes the proof that the height of a red-black tree is $O(log n)$ by maintaining the black height of the tree.

Source(s)
\begin{enumerate}
    \item \href{https://www.codesdope.com/course/data-structures-red-black-trees/}{Red-Black Trees : Properties, Black Height and Proof of its height}
    \item \href{https://pages.cs.wisc.edu/~siff/CS367/Notes/rbts.html}{CS 367-3 - Red-Black Trees - University of Wisconsin–Madison}
    \item \href{https://web.cse.ohio-state.edu/~lai.1/6331/0.Red-Black Trees.pdf}{Red-Black Trees - Department of Computer Science and Engineering}
    \item \href{https://www3.cs.stonybrook.edu/~skiena/214/lectures/lect25/lect25.html}{Red-Black Trees Lecture 25 - Stony Brook University}
\end{enumerate}

\textbf{User:} Could you give me a direct proof instead of a proof by induction?

\textbf{Bing Copilot:} Certainly! Here's a direct proof that the height of a red-black tree is $O(log n)$ by maintaining the black height of the tree:

The black height of a node is defined as the number of black nodes on any simple path from that node to a leaf node. The black height of all leaf nodes is zero. The black height of a red-black tree is the black height of its root node.

Let h be the black height of a red-black tree with n nodes. Since every path from the root to a leaf node contains at least h black nodes, the height of the tree is at most 2h. Therefore, the number of nodes in a red-black tree of height h is at most $2^{2h+1}-1$.

Since the height of a red-black tree is $O(log n)$, we can conclude that the number of nodes in a red-black tree of height h is at most $2^{2h+1}-1 = O(n)$.

This completes the direct proof that the height of a red-black tree is $O(log n)$ by maintaining the black height of the tree.

Source(s)
\begin{enumerate}
    \item \href{https://www.codesdope.com/course/data-structures-red-black-trees/}{Red-Black Trees : Properties, Black Height and Proof of its height}
    \item \href{https://www14.in.tum.de/lehre/2014WS/ea/split/sub-Red-Black-Trees-handout.pdf}{7.2 Red Black Trees Red Black Trees: Example - TUM}
    \item \href{https://pages.cs.wisc.edu/~siff/CS367/Notes/rbts.html}{CS 367-3 - Red-Black Trees - University of Wisconsin–Madison}
    \item \href{https://cs.gmu.edu/~rcarver/cs310/RedBlack.pdf}{Red-Back Trees - George Mason University}
\end{enumerate}

\textbf{User:} Isn't that proof a type of logical fallacy? You are using the claim as part of the proof.

\textbf{Bing Copilot:} You are correct. The proof I provided earlier is circular reasoning, which is a type of logical fallacy. I apologize for the mistake.

Here's a revised proof that does not use the claim as part of the proof:

The black height of a node is defined as the number of black nodes on any simple path from that node to a leaf node. The black height of all leaf nodes is zero. The black height of a red-black tree is the black height of its root node.

Let h be the black height of a red-black tree with n nodes. Since every path from the root to a leaf node contains at least h black nodes, the height of the tree is at most 2h. Therefore, the number of nodes in a red-black tree of height h is at most $2^{2h+1}-1$.

We can prove by induction that the minimum number of nodes in a red-black tree of height h is $2^{h+1}-1$. For the base case, consider a red-black tree of height 0. This tree has only one node, which is the root node. Therefore, the minimum number of nodes in the tree is 1, which is equal to $2^{0+1}-1$.

Now assume that the claim is true for all red-black trees of height less than h. Consider a red-black tree of height h. By the definition of a red-black tree, the root node must be black. Therefore, the two children of the root node must be red or black. If they are both black, then the black height of the tree is h-1, and the minimum number of nodes in the left and right subtrees is $2^h-1$ by the induction hypothesis. Therefore, the minimum number of nodes in the tree is $2^h-1 + 1 + 2^h-1 = 2^{h+1}-1$.

If the two children of the root node are both red, then the black height of the tree is h, and the minimum number of nodes in the left and right subtrees is $2^{h-1}-1$ by the induction hypothesis. Therefore, the minimum number of nodes in the tree is $2^{h-1}-1 + 1 + 2^{h-1}-1 + 1 = 2^{h+1}-1$.

In either case, the minimum number of nodes in the tree is $2^{h+1}-1$. Therefore, the height of the tree is at most $log(2^{2h+1}-1) = O(log n)$.

This completes the proof that the height of a red-black tree is $O(log n)$ by maintaining the black height of the tree.

Source(s)
\begin{enumerate}
    \item \href{https://www.codesdope.com/course/data-structures-red-black-trees/}{Red-Black Trees : Properties, Black Height and Proof of its height}
    \item \href{https://www14.in.tum.de/lehre/2014WS/ea/split/sub-Red-Black-Trees-handout.pdf}{7.2 Red Black Trees Red Black Trees: Example - TUM}
    \item \href{https://pages.cs.wisc.edu/~siff/CS367/Notes/rbts.html}{CS 367-3 - Red-Black Trees - University of Wisconsin–Madison}
    \item \href{https://cs.gmu.edu/~rcarver/cs310/RedBlack.pdf}{Red-Back Trees - George Mason University}
\end{enumerate}

\textbf{User:} Let's work on the part of the lecture dedicated to the insert and delete operations. For the insert operation, please give me its main motivation and pseudocode.

\textbf{Bing Copilot:} The main motivation behind using red-black trees is to ensure that the tree is always balanced, which guarantees that the time complexity for operations such as insertion, deletion, and searching is always O(log n), regardless of the initial shape of the tree. The color property of nodes in a red-black tree is used to ensure that the tree remains balanced.

Here's the pseudocode for inserting a new node into a red-black tree:
\begin{verbatim}
RB-Insert(T, z)
y = nil
x = root[T]
while x != nil
    y = x
    if key[z] < key[x]
        x = left[x]
    else
        x = right[x]
p[z] = y
if y == nil
    root[T] = z
else if key[z] < key[y]
    left[y] = z
else
    right[y] = z
left[z] = nil
right[z] = nil
color[z] = RED
RB-Insert-Fixup(T, z)
\end{verbatim}
This algorithm inserts a new node z into the red-black tree T. The new node is always inserted as a red node. After insertion of a new node, if the tree is violating the properties of the red-black tree, then we perform the following operations to maintain the balance of the tree:
\begin{verbatim}
RB-Insert-Fixup(T, z)
while color[p[z]] == RED
    if p[z] == left[p[p[z]]]
        y = right[p[p[z]]]
        if color[y] == RED
            color[p[z]] = BLACK
            color[y] = BLACK
            color[p[p[z]]] = RED
            z = p[p[z]]
        else
            if z == right[p[z]]
                z = p[z]
                LEFT-ROTATE(T, z)
            color[p[z]] = BLACK
            color[p[p[z]]] = RED
            RIGHT-ROTATE(T, p[p[z]])
    else
        (same as then clause with "right" and "left" exchanged)
color[root[T]] = BLACK
\end{verbatim}

This algorithm is used to maintain the properties of the red-black tree if insertion of a new node violates these properties. The algorithm uses rotations and color changes to ensure that the tree remains balanced.

Source(s)
\begin{enumerate}
    \item \href{https://www.programiz.com/dsa/insertion-in-a-red-black-tree}{Insertion in a Red-Black Tree - Programiz}
    \item \href{https://www.codesdope.com/course/data-structures-red-black-trees-insertion/}{Red Black Trees : Rotations and Insertions - CodesDope}
    \item \href{https://www.andrew.cmu.edu/user/mm6/95-771/examples/RedBlackTreeProject/dist/javadoc/redblacktreeproject/RedBlackTree.html}{Class RedBlackTree - andrew.cmu.edu}
    \item \href{https://www.eecs.umich.edu/courses/eecs380/ALG/red_black.html}{Data Structures and Algorithms: Red-Black Trees}
\end{enumerate}

\textbf{User:} Where did you get this pseudocode?

\textbf{Bing Copilot:} I apologize for the confusion earlier. The pseudocode I provided earlier was sourced from a website without proper citation. I have since found a reliable source for the pseudocode.

Here's the pseudocode for inserting a new node into a red-black tree, sourced from$^1$:
\begin{verbatim}
RB-Insert(T, z)
y = nil
x = root[T]
while x != nil
    y = x
    if key[z] < key[x]
        x = left[x]
    else
        x = right[x]
p[z] = y
if y == nil
    root[T] = z
else if key[z] < key[y]
    left[y] = z
else
    right[y] = z
left[z] = nil
right[z] = nil
color[z] = RED
RB-Insert-Fixup(T, z)
\end{verbatim}
This algorithm inserts a new node z into the red-black tree T. The new node is always inserted as a red node. After insertion of a new node, if the tree is violating the properties of the red-black tree, then we perform the following operations to maintain the balance of the tree:
\begin{verbatim}
RB-Insert-Fixup(T, z)
while color[p[z]] == RED
    if p[z] == left[p[p[z]]]
        y = right[p[p[z]]]
        if color[y] == RED
            color[p[z]] = BLACK
            color[y] = BLACK
            color[p[p[z]]] = RED
            z = p[p[z]]
        else
            if z == right[p[z]]
                z = p[z]
                LEFT-ROTATE(T, z)
            color[p[z]] = BLACK
            color[p[p[z]]] = RED
            RIGHT-ROTATE(T, p[p[z]])
    else
        (same as then clause with "right" and "left" exchanged)
color[root[T]] = BLACK
\end{verbatim}
This algorithm is used to maintain the properties of the red-black tree if insertion of a new node violates these properties. The algorithm uses rotations and color changes to ensure that the tree remains balanced.

Source(s)
\begin{enumerate}
    \item \href{https://www.programiz.com/dsa/insertion-in-a-red-black-tree}{Insertion in a Red-Black Tree - Programiz}
    \item \href{https://gcallah.github.io/algorithms/RedBlackTrees.html}{Design and Analysis of Algorithms: Red-Black Trees - GitHub Pages}
    \item \href{https://www.codingninjas.com/studio/library/insertion-in-red-black-trees}{Insertion In Red-Black Trees - Coding Ninjas}
    \item \href{https://iq.opengenus.org/red-black-tree-insertion/}{Red Black Tree: Insertion - OpenGenus IQ}
    \item \href{https://www.delftstack.com/howto/cpp/red-black-tree-cpp/}{Red Black Tree in C++ | Delft Stack}
\end{enumerate}

\textbf{User:} Please do the same for the delete operation.

\textbf{Bing Copilot:} Certainly! Here's the pseudocode for deleting a node from a red-black tree, sourced from$^1$:
\begin{verbatim}
RB-Delete(T, z)
if left[z] == nil or right[z] == nil
    y = z
else
    y = Tree-Successor(z)
if left[y] != nil
    x = left[y]
else
    x = right[y]
p[x] = p[y]
if p[y] == nil
    root[T] = x
else if y == left[p[y]]
    left[p[y]] = x
else
    right[p[y]] = x
if y != z
    key[z] = key[y]
    (copy y's satellite data into z)
if color[y] == BLACK
    RB-Delete-Fixup(T, x)
return y
\end{verbatim}
This algorithm deletes a node z from the red-black tree T. If the node to be deleted has no children, it is simply removed and the parent node is updated. If the node to be deleted has only one child, it is replaced with its child. If the node to be deleted has two children, then it is replaced with its in-order successor, which is the leftmost node in the right subtree. Then the in-order successor node is deleted as if it has at most one child. After the node is deleted, the red-black properties might be violated. To restore these properties, some color changes and rotations are performed on the nodes in the tree. The changes are similar to those performed during insertion, but with different conditions. The deletion operation in a red-black tree takes O(log n) time on average, making it a good choice for searching and deleting elements in large data sets$^1$.

Source(s)
\begin{enumerate}
    \item \href{https://www.geeksforgeeks.org/deletion-in-red-black-tree/}{Deletion in Red-Black Tree - GeeksforGeeks}
    \item \href{https://bing.com/search?q=red-black+trees+delete+operation+motivation+and+pseudocode}{Red Black Tree: Deletion - OpenGenus IQ}
    \item \href{https://www.gnu.org/software/mit-scheme/documentation/stable/mit-scheme-ref/Red_002dBlack-Trees.html}{Red-Black Trees (MIT/GNU Scheme 12.1)}
    \item \href{https://iq.opengenus.org/red-black-tree-deletion/}{Red Black Tree: Deletion - OpenGenus IQ}
    \item \href{https://medium.com/@snowme34/understand-red-black-tree-with-pseudo-codes-34aa31542e25}{Understand Red Black Tree with Pseudo-codes | by snowme34 - Medium}
\end{enumerate}

\textbf{User:} This is good. Please remind me of the class outline again.

\textbf{Bing Copilot:} Sure, here's the revised lecture outline for a 50-minute class on red-black trees:
\begin{enumerate}
    \item Introduction (5 minutes): Start the class with a brief recap of binary search trees and their limitations that were discussed in the last class. Introduce the concept of red-black trees as a solution to these problems.
    \item Understanding Red-Black Trees (10 minutes): Explain the properties of red-black trees and how they ensure balance in the tree. Discuss the color property of nodes and why it is essential.
    \item Analyzing Red-Black Trees (10 minutes): Use diagrams and examples to analyze the structure of red-black trees. Discuss the significance of black depth.
    \item Applying Operations on Red-Black Trees (15 minutes):
    \begin{enumerate}
        \item Insertion (7 minutes): Explain and demonstrate how to insert a node into a red-black tree. Discuss how the tree maintains its properties after insertion.
        \item Deletion (8 minutes): Explain and demonstrate how to delete a node from a red-black tree. Discuss how the tree maintains its properties after deletion.
    \end{enumerate}
    \item Summary and Q\&A (10 minutes): Summarize the key points of the lecture and open the floor for questions.
\end{enumerate}
This outline ensures that the class is interactive and gives students the opportunity to understand, analyze, and apply their knowledge of red-black trees. The Q\&A session at the end allows for clarification of doubts and deeper understanding.

\textbf{User:} Awesome. Let's wrap it up and give me the summary of the lecture.

\textbf{Bing Copilot:} Sure! Here's a summary of the lecture:
\begin{itemize}
    \item Red-black trees are a type of self-balancing binary search tree that use a specific set of rules to ensure that the tree is always balanced. This balance guarantees that the time complexity for operations such as insertion, deletion, and searching is always $O(log n)$, regardless of the initial shape of the tree.
    \item The color property of nodes in a red-black tree is used to ensure that the tree remains balanced. Each node in a red-black tree is either red or black. The color of a node is used to ensure that the tree remains balanced. The rules for coloring nodes are as follows:
    \begin{itemize}
        \item The root node must be black.
        \item Every leaf node (NULL node) must be black.
        \item If a node is red, its children must be black.
        \item Every path from a node to its descendant leaf nodes must contain the same number of black nodes.
    \end{itemize}
    \item Insertion and deletion operations on red-black trees are similar to those performed on binary search trees, but with additional color changes and rotations to ensure that the tree remains balanced.
    \item The deletion operation in a red-black tree takes $O(log n)$ time on average, making it a good choice for searching and deleting elements in large data sets.
\end{itemize}
I hope this summary helps!

\textbf{User:} Please compile the learning objectives, outline, and content into a single large lecture document.

\textbf{Bing Copilot:} Sure! Here's a compiled lecture document on red-black trees:

\textbf{Red-Black Trees}

\textbf{Learning Objectives}

By the end of this lecture, students should be able to:
\begin{enumerate}
    \item Understand the concept of red-black trees: Students should be able to explain what red-black trees are, their properties, and why they are used in computer science.
    \item Analyze the structure of red-black trees: Students should be able to dissect a given red-black tree, identify its components, and explain the role of each component.
    \item Apply the operations of red-black trees: Students should be able to perform basic operations on red-black trees such as insertion, deletion, and searching.
    \item Evaluate the performance of red-black trees: Students should be able to assess the time complexity of operations on red-black trees and compare it with other data structures.
    \item Create red-black trees: Given a set of data, students should be able to construct a red-black tree.
\end{enumerate}

\textbf{Lecture Outline}
\begin{enumerate}
    \item Introduction (5 minutes): Start the class with a brief recap of binary search trees and their limitations that were discussed in the last class. Introduce the concept of red-black trees as a solution to these problems.
    \item Understanding Red-Black Trees (10 minutes): Explain the properties of red-black trees and how they ensure balance in the tree. Discuss the color property of nodes and why it is essential.
    \item Analyzing Red-Black Trees (10 minutes): Use diagrams and examples to analyze the structure of red-black trees. Discuss the significance of black depth.
    \item Applying Operations on Red-Black Trees (15 minutes):
    \begin{itemize}
        \item Insertion (7 minutes): Explain and demonstrate how to insert a node into a red-black tree. Discuss how the tree maintains its properties after insertion.
        \item Deletion (8 minutes): Explain and demonstrate how to delete a node from a red-black tree. Discuss how the tree maintains its properties after deletion.
    \end{itemize}
    \item Summary and Q\&A (10 minutes): Summarize the key points of the lecture and open the floor for questions.
\end{enumerate}

\textbf{Lecture Content}

\textbf{Introduction}
A binary search tree (BST) is a node-based binary tree data structure that has the following properties: The left subtree of a node contains only nodes with keys lesser than the node’s key. The right subtree of a node contains only nodes with keys greater than the node’s key. The left and right subtree each must also be a binary search tree . Binary search trees are used to store data in a way that allows for efficient searching, insertion, and deletion of elements. They are commonly used in computer science and are the foundation for many other data structures.

However, binary search trees have some limitations. If the tree is not balanced, the time complexity for operations such as insertion, deletion, and searching can be O(n), where n is the number of nodes in the tree. This can be a problem when dealing with large data sets. To overcome this limitation, we use red-black trees.

Red-black trees are a type of self-balancing binary search tree that use a specific set of rules to ensure that the tree is always balanced. This balance guarantees that the time complexity for operations such as insertion, deletion, and searching is always $O(log n)$, regardless of the initial shape of the tree.

\textbf{Understanding Red-Black Trees}

Red-black trees are similar to binary search trees, but with an additional color property for each node. Each node in a red-black tree is either red or black. The color of a node is used to ensure that the tree remains balanced. The rules for coloring nodes are as follows:
\begin{itemize}
    \item The root node must be black.
    \item Every leaf node (NULL node) must be black.
    \item If a node is red, its children must be black.
    \item Every path from a node to its descendant leaf nodes must contain the same number of black nodes.
\end{itemize}
These rules ensure that the longest path from the root to a leaf node is no more than twice as long as the shortest path . This balance guarantees that the time complexity for operations such as insertion, deletion, and searching is always $O(log n)$, regardless of the initial shape of the tree . Red-black trees are used in many applications, including memory allocation algorithms, compilers, and databases .

\textbf{Analyzing Red-Black Trees}
The structure of a red-black tree can be analyzed using diagrams and examples. The black depth of a node is defined as the number of black nodes on any simple path from that node to a leaf node. The black depth of all leaf nodes is zero. The black depth of a red-black tree is the black depth of its root node.

The significance of black depth is that it ensures that the longest path from the root to a leaf node is no more than twice as long as the shortest path . This balance guarantees that the time complexity for

\textit{Author note: it cut off the response here for some unknown reason.}

\textbf{User:} Please finish

\textbf{Bing Copilot:} I apologize for the incomplete response earlier. Here's the rest of the lecture content:

\textbf{Applying Operations on Red-Black Trees}

Insertion and deletion operations on red-black trees are similar to those performed on binary search trees, but with additional color changes and rotations to ensure that the tree remains balanced.

Here's the pseudocode for inserting a new node into a red-black tree:
\begin{verbatim}
RB-Insert(T, z)
y = nil
x = root[T]
while x != nil
    y = x
    if key[z] < key[x]
        x = left[x]
    else
        x = right[x]
p[z] = y
if y == nil
    root[T] = z
else if key[z] < key[y]
    left[y] = z
else
    right[y] = z
left[z] = nil
right[z] = nil
color[z] = RED
RB-Insert-Fixup(T, z)
\end{verbatim}
This algorithm inserts a new node z into the red-black tree T. The new node is always inserted as a red node. After insertion of a new node, if the tree is violating the properties of the red-black tree, then we perform the following operations to maintain the balance of the tree:
\begin{verbatim}
RB-Insert-Fixup(T, z)
while color[p[z]] == RED
    if p[z] == left[p[p[z]]]
        y = right[p[p[z]]]
        if color[y] == RED
            color[p[z]] = BLACK
            color[y] = BLACK
            color[p[p[z]]] = RED
            z = p[p[z]]
        else
            if z == right[p[z]]
                z = p[z]
                LEFT-ROTATE(T, z)
            color[p[z]] = BLACK
            color[p[p[z]]] = RED
            RIGHT-ROTATE(T, p[p[z]])
    else
        (same as then clause with "right" and "left" exchanged)
color[root[T]] = BLACK
\end{verbatim}
This algorithm is used to maintain the properties of the red-black tree if insertion of a new node violates these properties. The algorithm uses rotations and color changes to ensure that the tree remains balanced.

Here's the pseudocode for deleting a node from a red-black tree:
\begin{verbatim}
RB-Delete(T, z)
if left[z] == nil or right[z] == nil
    y = z
else
    y = Tree-Successor(z)
if left[y] != nil
    x = left[y]
else
    x = right[y]
p[x] = p[y]
if p[y] == nil
    root[T] = x
else if y == left[p[y]]
    left[p[y]] = x
else
    right[p[y]] = x
if y != z
    key[z] = key[y]
    (copy y's satellite data into z)
if color[y] == BLACK
    RB-Delete-Fixup(T, x)
return y
\end{verbatim}
This algorithm deletes a node z from the red-black tree T. If the node to be deleted has no children, it is simply removed and the parent node is updated. If the node to be deleted has only one child, it is replaced with its child. If the node to be deleted has two children, then it is replaced with its in-order successor, which is the leftmost node in the right subtree. Then the in-order successor node is deleted as if it has at most one child. After the node is deleted, the red-black properties might be violated. To restore these properties, some color changes and rotations are performed on the nodes in the tree. The changes are similar to those performed during insertion, but with different conditions. The deletion operation in a red-black tree takes $O(log n)$ time on average, making it a good choice for searching and deleting elements in large data sets.

\textbf{Summary}
\begin{itemize}
    \item Red-black trees are a type of self-balancing binary search tree that use a specific set of rules to ensure that the tree is always balanced. This balance guarantees that the time complexity for operations such as insertion, deletion, and searching is always $O(log n)$, regardless of the initial shape of the tree.
    \item The color property of nodes in a red-black tree is used to ensure that the tree remains balanced. Each node in a red-black tree is either red or black. The color of a node is used to ensure that the tree remains balanced. The rules for coloring nodes are as follows:
    \begin{itemize}
    \item The root node must be black.
    \item Every leaf node (NULL node) must be black.
    \item If a node is red, its children must be black.
    \item Every path from a node to its descendant leaf nodes must contain the same number of black nodes.
    \end{itemize}
    \item Insertion and deletion operations on red-black trees are similar to those performed on binary search trees, but with additional color changes and rotations to ensure
\end{itemize}







\end{document}